\documentclass[aps,prd,twocolumn,nofootinbib,floatfix]{revtex4-1}

\usepackage{color}
\usepackage{amsmath}
\usepackage{amssymb}
\usepackage[english]{babel}
\usepackage{graphicx}
\usepackage{subfigure}
\usepackage{graphicx}
\usepackage{float}

\usepackage{bclogo}

\newcommand{\be}{\begin{equation}}
\newcommand{\ee}{\end{equation}}
\newcommand{\bea}{\begin{eqnarray}}
\newcommand{\eea}{\end{eqnarray}}

\newcommand{\bit}{\begin{itemize}}
\newcommand{\eit}{\end{itemize}}

\newcommand{\pade}[2]{\begin{small}$[ \, #1 \, | \, #2 \, ]$\end{small}}
\newcommand{\padelog}[4]{\begin{small}$[ \, #1 \, | \, #2 \, ; \, #3 \, | \, #4 \, ]$\end{small}}

\usepackage{verbatim}

\extrafloats{1000}

\begin{document}

\title{The analytic structure of the lattice Landau gauge gluon and ghost propagators}

\author{Alexandre F. Falc\~ao}
\email{alfa@uc.pt}
\author{Orlando Oliveira}
\email{orlando@uc.pt}
\author{Paulo J. Silva}
\email{psilva@uc.pt}
\affiliation{CFisUC, Department of Physics, University of Coimbra, 3004-516 Coimbra, Portugal}

\begin{abstract}
Starting from the lattice Landau gauge gluon and ghost propagator data we use a sequence of Pad\'e approximants, 
identify the poles and zeros for each approximant and map them into the analytic structure of the propagators. 
For the Landau gauge gluon propagator the Pad\'e analysis identifies a pair of complex conjugate poles and a branch cut along the negative 
real axis of the Euclidean $p^2$ momenta. 
For the Landau gauge ghost propagator the Pad\'e analysis shows a single pole at $p^2 = 0$ and a branch cut also along the negative real axis of 
the Euclidean $p^2$ momenta.  
The method gives precise estimates for the gluon complex poles, that agree well with other estimates  found in the literature.
For the branch cut the Pad\'e analysis gives, at least, a rough estimate of the corresponding branch point.
\end{abstract}

\maketitle

\section{Introduction and Motivation}

Quantum Chromodynamics (QCD) is a non-abelian gauge theory associated with the SU(3) color group that describes the interactions between quarks 
and gluons \cite{Alkofer:2000wg,Fischer:2006ub,Binosi:2009qm}.
Its fundamental quanta have never been observed in an experiment \cite{Perl:2009zz,Tanabashi:2018oca}. 
This negative result suggests that the single particle states associated with quarks and gluons 
do not belong to the Hilbert space of the physical states. Thus, quarks and gluons can only exist as components of the physical states, identified as the color singlet
states, a statement that is normally phrased saying that quarks and gluons are confined particles.
Making the bridge between the underlying quark and gluon dynamics to the observed particle states is far from trivial and it certainly requires solving QCD
beyond its perturbative solution. Confinement is not the only hadronic property that calls for a non-perturbative solution of QCD. 
In general the understanding of hadronic phenomena, as for example the realisation of the chiral symmetry breaking mechanism, 
calls for solutions outside the perturbative approach to QCD.

In a quantum field theory as  QCD, the dynamical information is summarised in its correlation functions. 
The quark, the gluon and the ghost propagators are among the simplest Green's functions that can be considered 
and, together with a finite number of vertices, are the essential building blocks required to understand hadrons \cite{Huber:2018ned}. 
They contain information on the physical spectra, on the dynamical properties that experimentally are seen as form factors 
and, at finite temperature and/or density, the correlation functions encode the transport properties. Furthermore, the propagators are 
necessary for the computation  of the hadronic phase diagram.  The two point correlation functions also contain information on confinement,
on the chiral symmetry breaking mechanism and on the generation of mass scales that are associated with its fundamental fields. 
These infrared mass scales regularise the theory at low energies.
The knowledge of the pole structure of the propagators and the position of their branch cuts, i.e. of their analytic structure, is relevant to access many
hadronic properties and to the understanding of non-perturbative phenomena as e.g. confinement and chiral symmetry breaking at a fundamental level.

Most non-perturbative approaches to quantum field theory rely on the Euclidean formulation of the theory. However,
if one uses the Euclidean formulation 
the observables or quantities that are associated with time-like momenta are not easily accessible.
In general, by doing the analytic continuation of the Euclidean correlation functions, i.e. the Schwinger functions, it is possible to get
the corresponding Minkowski space Green functions, the Wightman functions. 
This can only be achieved  if the analytic structure of the Green functions is known in advance.

In perturbation theory the analytical continuation from Minkowski to Euclidean space is done via the usual Wick rotation \cite{Ramond:1981pw}. 
However, beyond perturbation
theory there is no clear rule to analytically  continue the Schwinger functions to complex momenta.  
For example, there are indications that the propagators can have complex poles
\cite{Krein:1990sf,Maris:1995ns,Dudal:2010tf,Cucchieri:2011ig}. 
The presence of complex poles in the Argand plane make the usual Wick rotation impractical but not the analytical extension of the correlation functions
\cite{Eichmann:2019dts}.
It has also been argued by some authors that the use of integral representations can solve the problem of accessing Minkowski space correlation functions 
from the corresponding Euclidean functions \cite{Kusaka:1995za,Castro:2019tlh}. However, it still remains to be shown that this achievement works. 
Certainly, the precise determination of the structure of cuts and poles of the propagator for complex $p^2$ is, by itself, a fundamental problem in physics and also
a non-trivial mathematical problem.  

Herein, we make an attempt to access the analytic structure of the Landau gauge gluon and ghost propagators for pure Yang-Mills theory, taken from lattice QCD 
simulations, using sequences of Pad\'e approximants.
The use of Pad\'e approximants in Physics is common and used to address many problems. A far from complete list of examples can be found 
in \cite{Schlessinger:1966zz,Schlessinger68,Basdevant72} and references therein. 
Indeed, the Pad\'e approximants lies at the heart of investigations on the analytic structure of physical quantities 
\cite{VidbergSerene77,SanzCillero:2010mp,Queralt:2010sv,Boito:2018rwt,TripoldSmekal19,Binosi:2019ecz}
or on the identification of singularities for several types of functions \cite{Hillion77,Billi94,YamadaIkeda2014}.

In what concerns the QCD propagators, in \cite{Stingl:1994nk} a general scheme based on Pad\'e approximants to solve the Dyson-Schwinger equations 
was suggested but, to the best knowledge of the authors, it was never implemented or tried. 
There have been attempts to determine the analytic structure of the propagators from the Dyson-Schwinger solutions for the propagators
\cite{Alkofer:2003jj,Alkofer:2004cw} relying on the computation of the Schwinger functions, combined with the use of  functions that are able to 
reproduce the non-perturbative solutions of the theory and also well known features of theory on the ultraviolet regimen.
In \cite{Strauss:2012dg} there was a tentative to solve the (approximate) Dyson-Schwinger equations 
for the gluon and ghost propagators in pure QCD  for complex $p^2$ directly. 
The tree level solution for the propagators from the Gribov-Zwanziger \cite{Gribov:1977wm,Dudal:2007cw,Dudal:2008sp,Vandersickel:2012tz,Capri:2017bfd} 
class of actions is a ratio of polynomials and, therefore, can be seen as Pad\'e approximants to the propagators. 
As described in \cite{Dudal:2010tf,Cucchieri:2011ig,Oliveira:2012eh,Dudal:2018cli}, these type of functional form describe extremely well the 
Landau gauge lattice gluon propagator data.

The study of the analytic structure of quantum field theories using the Dyson-Schwinger equations is not restricted to QCD and, for example,
in \cite{Krein:1990sf,Maris:1995ns} the analytic structure of other types of theories was also considered. 
Also, in \cite{Dudal:2013yva,Oliveira:2016stx,Dudal:2019gvn} there has been a tentative to identify the branch cut for the gluon and ghost propagators relying
on its K\"allen-Lehmann representation by measuring directly, from the lattice data, its spectral function at zero temperature. 
All these studies suggest that the gluon and ghost propagators have a non-trivial analytic structure that requires to be understood.

This paper is organised as follows. In Sec. \ref{Sec:Pade} we review the fundamentals of Pad\'e approximants, set the notation and discuss its applications to
some test functions.
In Sec. \ref{Sec:general} we look at the quality of the lattice data for the gluon propagator to check for the presence of logarithmic behaviour in the lattice
data and discuss the class of approximants to be used to describe the lattice propagator.
In Sec. \ref{Sec:gluon} the Pad\'e analysis is performed for the gluon propagator and in Sec. \ref{Sec:ghost} we report on the results for the ghost
propagator.
Finally, in Sec. \ref{Sec:ultima} we summarise our results, discuss its meaning and look for future work.

\section{Elements of  Pad\'e approximants \label{Sec:Pade}}

The idea behind the Pad\'e approximants is to represent a given function by a ratio of polynomials, not necessarily of the same degree. 
By approximating a function by a ratio of polynomials, a set of zeros and poles is associated to the each Pad\'e approximant.
However, not all the zeros and  poles are meaningful. In general, changing the degree of the polynomials changes the position of the
 zeros and poles of the approximants. Still, there is a subset of zeros and poles whose position in the complex plane remains stable, i.e. it does not
 depend on the Pad\'e approximant used. 
It is these stable set of zeros and poles that can  be associated with the analytic structure and, thererore, with 
physical properties. All the remaining zeros and poles are artefacts of the 
approximation. 

The stable poles and zeros are the remnants of the analytic structure of the original function. Some can be identified with
single poles, while others are certainly representations of more complex structures as multiple poles or branch cuts.
In particular, a branch cut can be identified as a sequence of sets of close zeros and poles whose position in the complex plan is essentially
independent of the Pad\'e approximant used.

For a given propagator $D(p^2)$ its \pade{M}{N} Pad\'e approximant is defined as
\begin{equation}
   D(p^2) \approx P^M_N(p^2) = \frac{Q_M(p^2)}{R_N(p^2)} \ ,
   \label{Eq:usual_Pade_M_N_0}
\end{equation}
where
\begin{eqnarray}
     Q_M(p^2) & = & q_0 + \cdots + q_M \, \left( p^2 \right)^M \ , \\
     R_N(p^2) & = & 1 + \cdots + r_N \, \left( p^2 \right)^N \ . 
\end{eqnarray} 
In our convention, the coefficient of the lowest order term in the polynomial at the denominator is set to one.   

A fundamental result that gives support to the use of Pad\'e approximants is Pommerenke’s Theorem \cite{Pomm73}. It states that for a meromorphic function
$f(z)$,  the Pad\'e sequences \pade{M}{M+k}, with fixed $k$, converge to $f(z)$ in any compact set of the complex plane. 
In the Pad\'e approximant, single poles of $f(z)$, are sets of zero area, and appear in the \pade{M}{N} approximants as stable poles for sufficiently 
large values of $M$. 
The Pad\'e approximants have also poles whose position depends strongly on $M$ and $N$, or appear with nearby zeros that define the so called 
Froissart doublets \cite{Baker75,BaMo96,BeOr99,SanzCillero:2010mp,Queralt:2010sv}. 
The absolute value of the residua of these Froissart doublets is small due to the nearby zeros. Moreover, these doublets appear at sufficiently large values of
$M$ and $N$ and are artefacts associated with the use of ratio of polynomials.
For practical purposes, it appears that the preferable Pad\'e approximants are diagonal, i.e. are of the form \pade{M}{M}, 
or are nearby diagonal sequences where $k = \pm 1$.

For certain classes of functions, the convergence of the Padé sequences to the right limit can be proved explicitly.  
Among this class of functions are those of the Stieltjes type whose general structure is represented by
\begin{equation}
   f(z) = \int^{+\infty}_0 ~\frac{ d \mu (t) }{ 1 + z \, t} \  ,  ~  \qquad |Arg(z)|  \, < \, \pi
   \label{Eq:Stieltjies}
\end{equation}
where $\mu(t)$ is a measure defined in $t \in [0, \, + \infty [$. 
The  Kall\"en-Lehmann integral representation for the propagators of physical particles belongs to the class of Stieltjes functions. 
However, for the gluon and ghost, which are confined particles, the corresponding propagators do not have necessarily an integral representation of the 
type given in Eq. (\ref{Eq:Stieltjies}).  The numerical experiments performed in \cite{Dudal:2013yva,Oliveira:2016stx,Dudal:2019gvn,Binosi:2019ecz} 
show that it is possible to build an integral representation for the propagators if $\mu (t)$ is no longer a measure in $[0, \, + \infty [$ 
or when the integration range is extended.
This is no proof that the Pad\'e approximants sequences work well for the gluon and ghost propagators but given its general properties,
given the predictive power associated with the Pad\'e approximants in many situations, 
it seems reasonable to explore the  use of sequences of Pad\'e approximants to investigate the analytic structure of the propagators.

The traditional definition of the Pad\'e approximants and, in particular, the computation of the polynomial coefficients rely on the ability to perform
series expansions that, for the lattice propagators, are not possible. Therefore,  the numerical experiments to be reported in this manuscript
rely in the determination of the absolute minimum of an objective function, the reduced $\chi^2$ for the corresponding problem, to determine the coefficients
of the polynomials. The value of the reduced $\chi^2$ at the minimum will also describe the quality of the approximation achieved with the approximant. 

For our definition of the Pad\'e approximant it implies solving a non-linear global optimisation problem. 
The computation of the absolute minimum of a non-linear function does not have, in general, a solution. 
For the numerical experiments, we rely on the global optimisation methods available within \textit{Mathematica} 
\cite{Math} software package. Namely, we rely on their implementation of the differential evolution (DE) method and of the simulated
annealing (SA) method, two standard numerical methods that address the determination of the absolute extreme of a generic function.

\subsection{Numerical tests with Padé approximants on test functions \label{Sec:TestFunc}}

A first flavour on an analysis of a sequence of Pad\'e approximants can be obtained looking at simple functions that
are somehow related to the QCD propagators. This is the motivation to study
\begin{eqnarray}
   D_1(p^2) & = & \frac{1}{p^2 + m^2} \ ,     \label{Eq:TestFuncs1} \\
   D_2(p^2)   & = & \log (p^2 + m^2) \  ,   \label{Eq:TestFuncs2}  \\
    D_3(p^2)  & = & \frac{1}{p^2} \left( \omega \, \log p^2 + 1 \right)^\gamma  \label{Eq:TestFuncs3}
\end{eqnarray} 
that are inspired in the perturbative solution of QCD for the propagators. 
The function $D_1(p^2)$ is the tree level expression for the propagator and has a simple pole at $p^2 = - m^2$. 
The function $D_2(p^2)$ has a branch cut and will allow to understand how a branch cut appears in a sequence of Pad\'e approximants analysis.
Finally, the function $D_3(p^2)$ reproduces the expected behaviour for the propagator in the ultraviolet regime and has a simple pole and a branch cut.
In the  analysis of the test functions we have also considered other variants than those reported here. However, the results for $D_1(p^2)$ to 
$D_3(p^2)$ illustrate well the outcome of all the trials.

\begin{figure}[t] 
   \centering
   \includegraphics[width=3in]{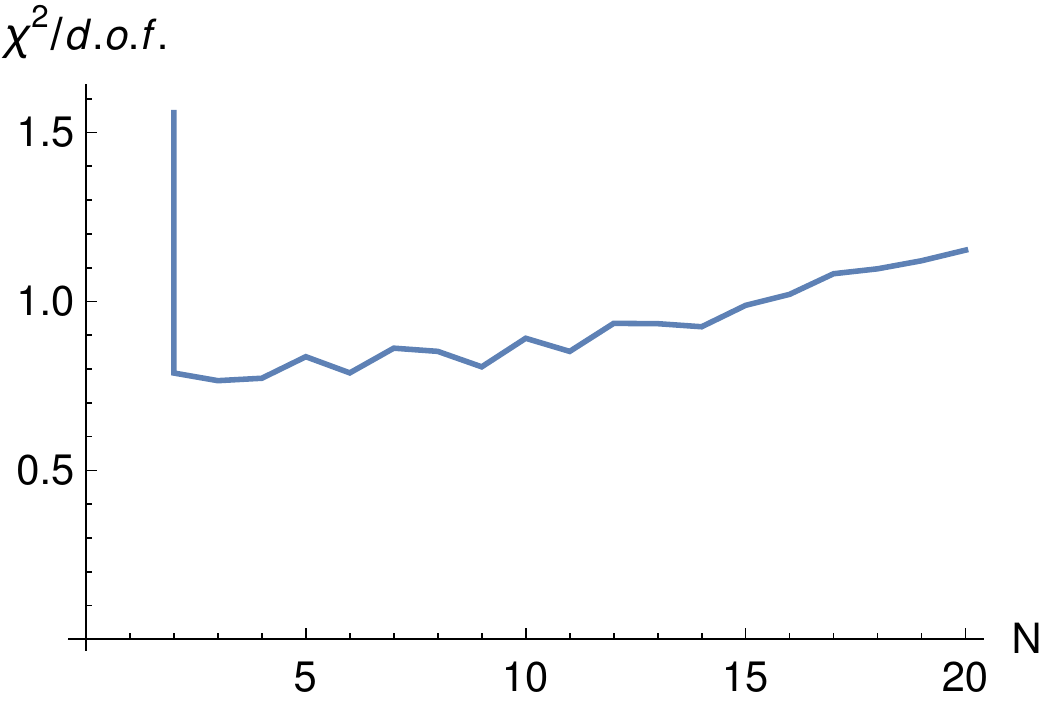}  \\
   \includegraphics[width=3in]{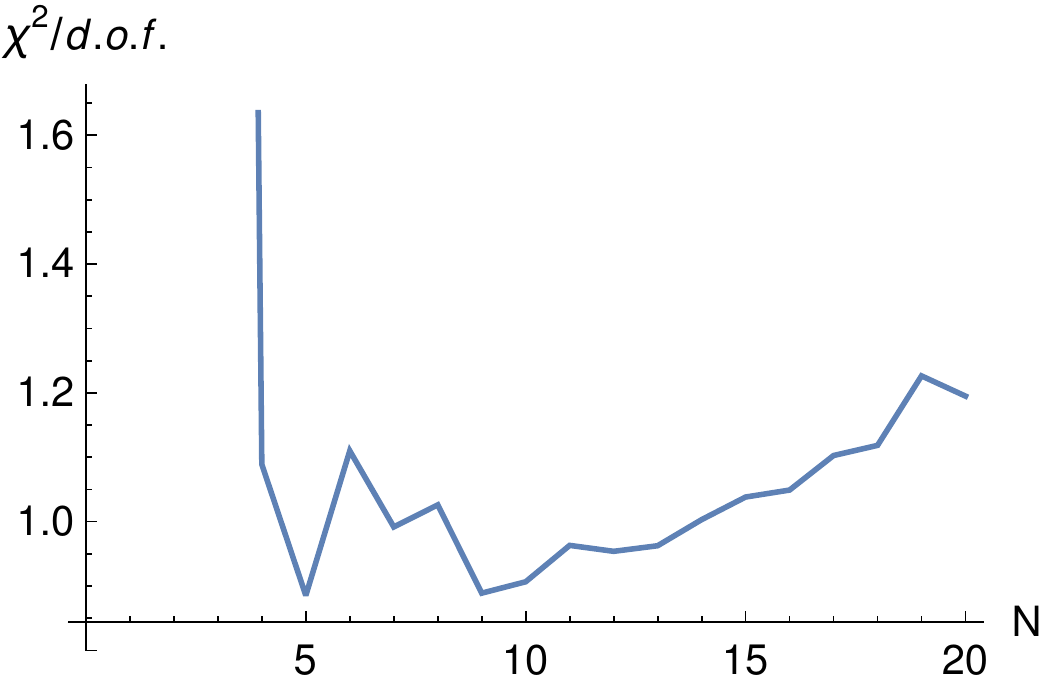} \\
   \includegraphics[width=3in]{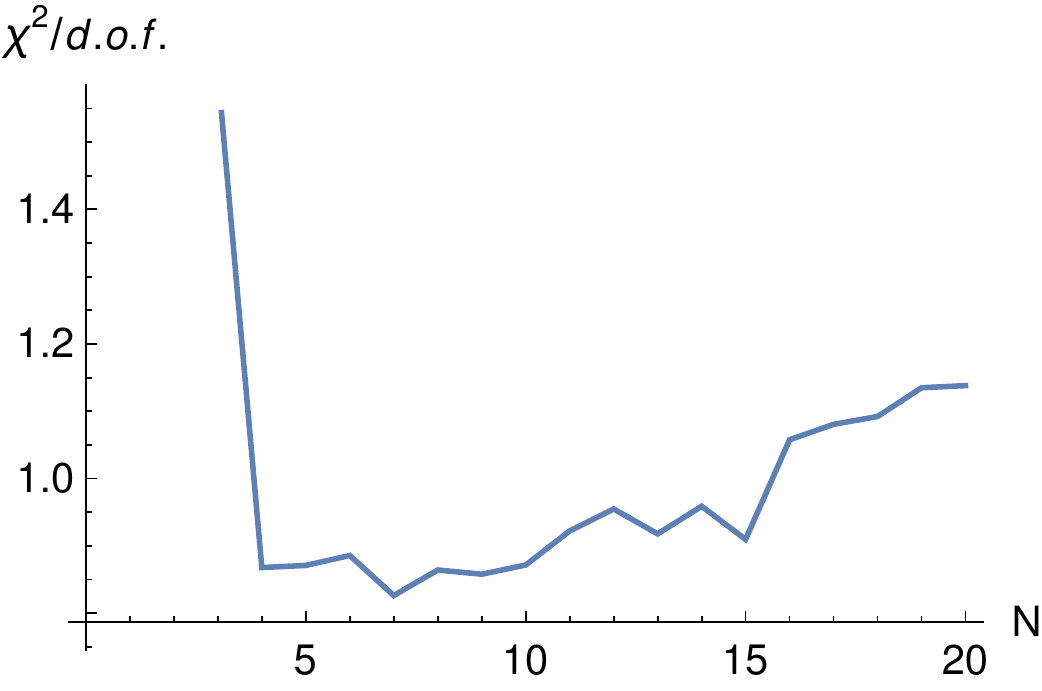} 
   \caption{The $\chi^2/d.o.f.$ obtained by minimising the $\chi^2$ defined in Eq. (\ref{Eq:objectivefunction}) for the functions (\ref{Eq:TestFuncs1}) (top), 
                 (\ref{Eq:TestFuncs2}) using for both functions $m^2 = 0$ (middle) and (\ref{Eq:TestFuncs3}) with $\omega = 0.3$  and $\gamma = -13/22$ (bottom). 
                 The values reported are obtained with the DE method for $\epsilon = 1\%$. Similar curves can be made for different values of $\epsilon$,
                 number of momentum data points and for the SA algorithm.}
   \label{fig:Test_Chi2}
\end{figure}

In the current section $p^2$ is dimensionless and, to simulate the analysis of the lattice data, instead of using directly the analytical functions $D_1(p^2)$
to $D_3(p^2)$, a set of uniformly random distributed $p$ in the range $p \in [ 0, \, 8]$ was generated. 
These ``lattice data'' is not the direct result of using the above analytical forms but, instead, we take $D(p^2) ( 1 + \epsilon \, \mathcal{N} ( 0, \sigma) )$, where 
$ \mathcal{N} ( 0, \sigma) $ is a normal distribution with mean value zero and width $\sigma = 1$, 
with an associated error that is given by $\epsilon ~ D(p^2) $. 
In the numerical experiments for the functions (\ref{Eq:TestFuncs1}) - (\ref{Eq:TestFuncs3}) we set
$\epsilon = 1 \%$ and $0.1\%$ and considered 100 data points for $p$. 
The lattice data for the gluon and the ghost propagators used below has more than a hundred data points, with statistical errors that are
within the same ballpark. The analysis briefly reported here for the test functions is a less favourable situation compared with the real data
and, in this sense, it provides a worst case scenario. 
We have also done the analysis of the test functions considering more data points and the numerical experiments show that by increasing the 
number of data points, the results of the Pad\'e analysis become closer to the original functions.

\begin{figure}[t] 
   \centering
   \includegraphics[width=3in]{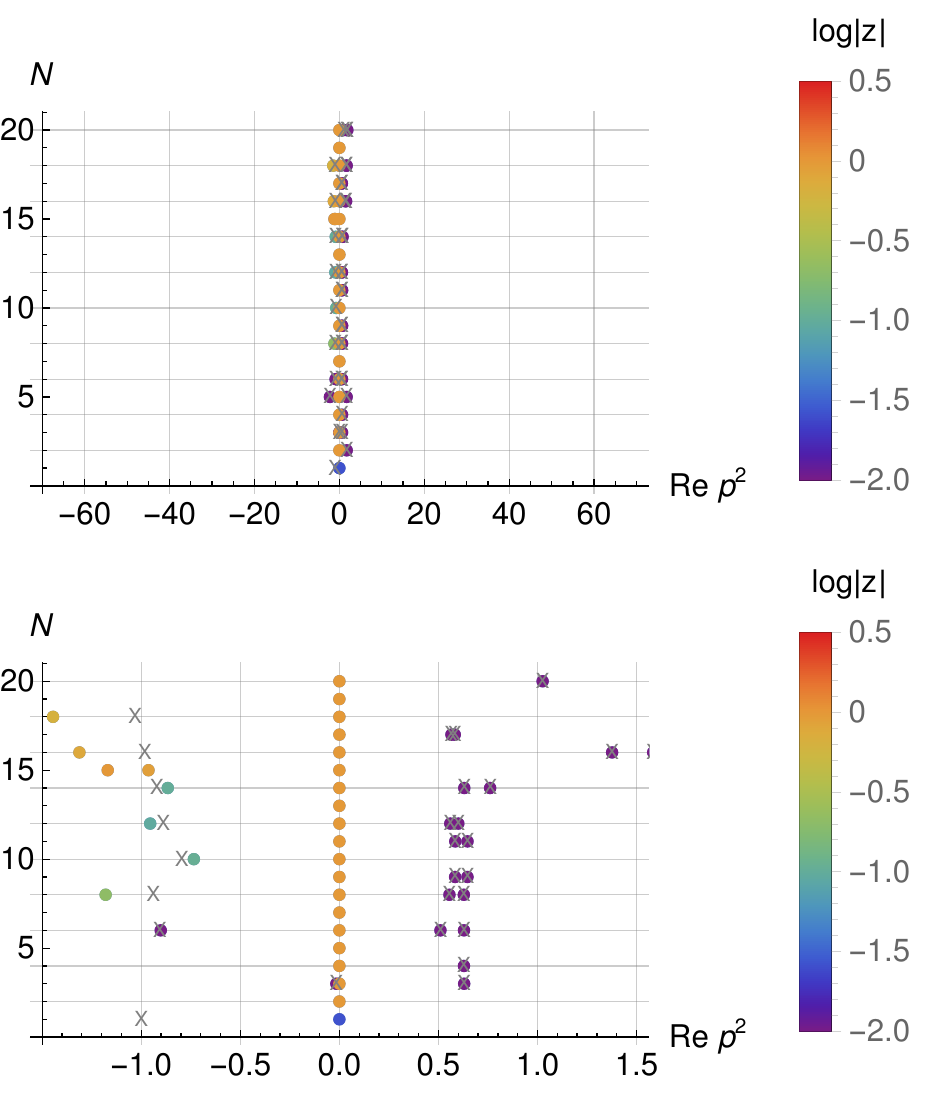}  \\
   \includegraphics[width=3in]{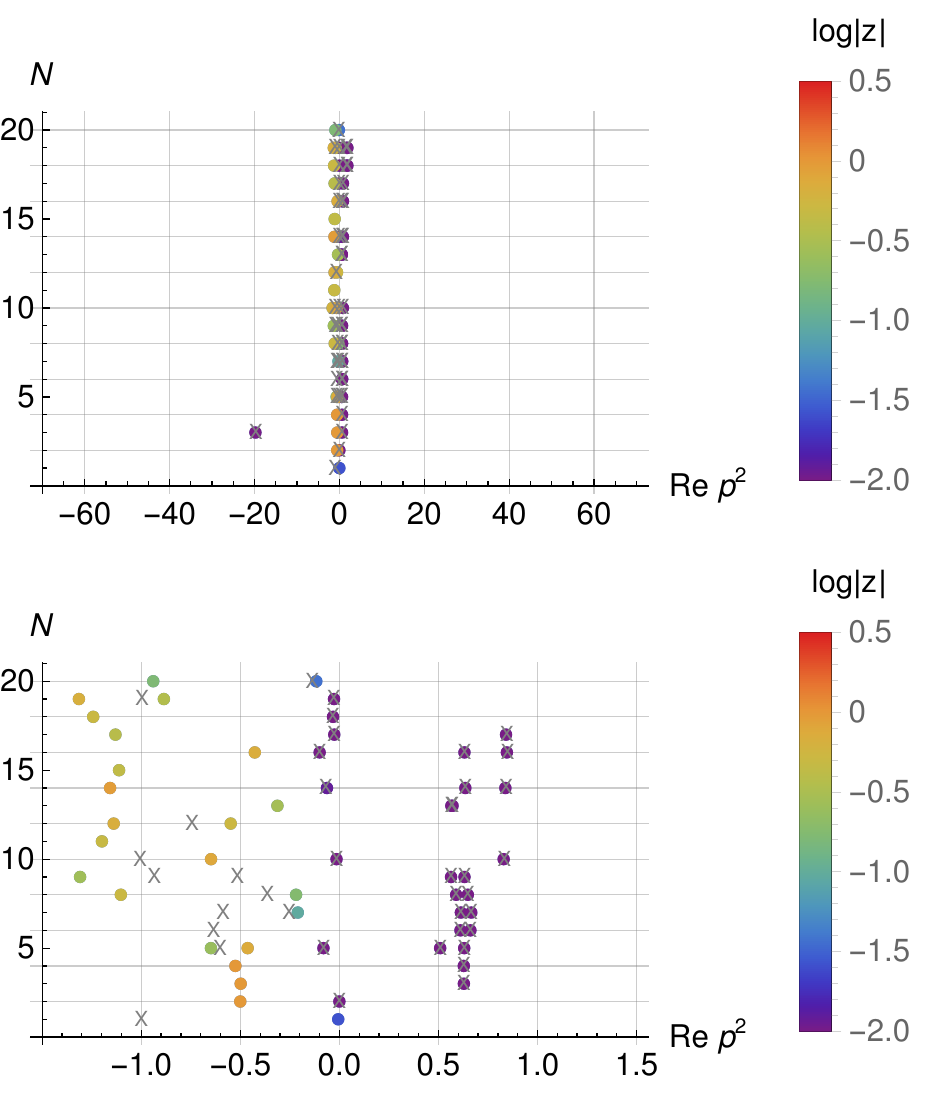} 
   \caption{Distribution of zeros (crosses) and poles (circles) for on-axis momenta as a function of $N$, resulting from the Padé analysis for the data generated 
                  with Eq. (\ref{Eq:TestFuncs1}) for $m^2 = 0$ (top two plots)  and for  $m^2 = 0.5$ (bottom two plots). The scale on the left refers to the absolute 
                  values of the residua.}
   \label{fig:Test_OnAxis_tree}
\end{figure}

\begin{figure}[t] 
   \centering
   \includegraphics[width=3.4in]{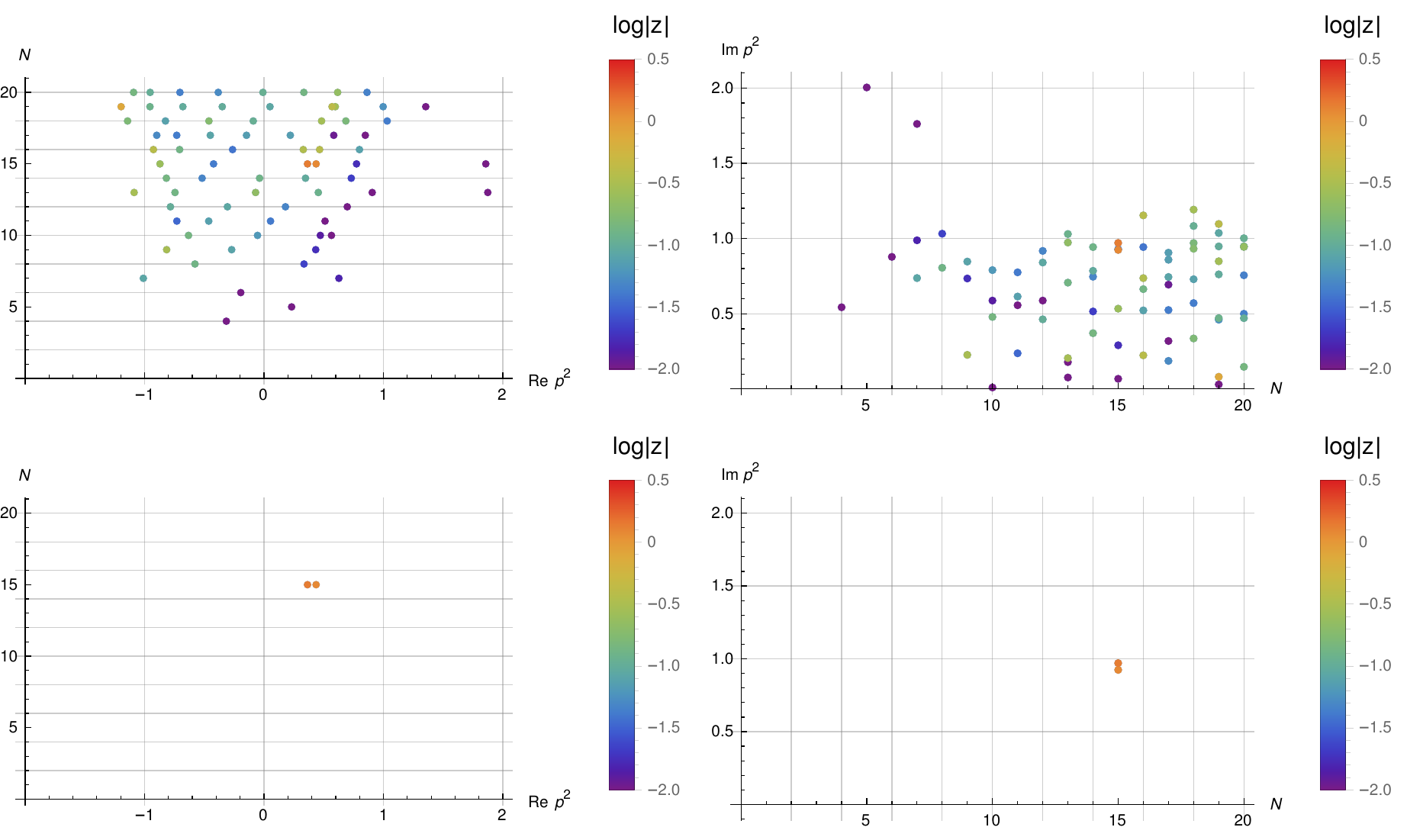} 
   \caption{Distribution of poles for complex $p^2$ as a function of $N$, resulting from the Padé analysis for the data generated with Eq. (\ref{Eq:TestFuncs1}) for 
                 $m^2 = 0$. On the top plot are all the poles with the legend showing the corresponding residua, while on the bottom plot the poles whose residua
                 $|A|$ is smaller than one are omitted.}
   \label{fig:Test_OffAxis_tree}
\end{figure}

The coefficients of the polynomials for the Pad\'e approximants are computed minimising the objective function defined as the  $\chi^2$ that takes
into account the errors on the data, i.e.
\begin{equation}
   \chi^2 = \sum_{j=1}^N \left( \frac{ D(p^2_j) - D_{Lat}(p^2_j) }{ \sigma(p^2_j) } \right)^2 \ ,
   \label{Eq:objectivefunction}
\end{equation}
where the sum is over the data points, $D(p^2) = P^M_N(p^2)$, $D_{Lat}(p^2)$ are the data points for the given function and $\sigma(p^2)$ are the associated statistical
error with $D_{Lat}(p^2)$. Our analysis does not takes into account the correlations between the various momenta.

The coefficients of the polynomials are defined by estimating the absolute minima of $\chi^2$
with  the routines for global optimisation included in \textit{Mathematica} \cite{Math}.

In general, the results for the zeros and poles obtained with the DE and SA methods have similar patterns, with possible deviations in the detail.
Further, for the numerical experiments associated to the function given in Eqs. (\ref{Eq:TestFuncs1}) to  (\ref{Eq:TestFuncs3}) 
only the Pad\'e approximants of type \pade{N-1}{N} and up to $N = 20$ were considered. 
The values obtained for the $\chi^2 / d.o.f.$ at the global extrema are, for all functions and for the two methods considered, in the range 0.8 - 1.2.
In Fig. \ref{fig:Test_Chi2} we show an example of the reduced $\chi^2$ obtained with the DE method. Similar curves can be drawn for the SA method.
It is reassuring  that both methods return very close values for the $\chi^2/d.o.f.$  In this section we will show, preferably, the results obtained with the
DE method. Moreover, given that this section aims to illustrate the performance of the Pad\'e analysis on the test functions, only a selected set of plots
will be considered.

In Fig. \ref{fig:Test_OnAxis_tree} we report how a single pole can be identified by a sequence of Pad\'e approximants 
associated with the function given by Eq. (\ref{Eq:TestFuncs1}) for $m^2 = 0$ (top two plots) and for $m^2 = 0.5$ (bottom two plots). 
As seen, the Pad\'e sequence
reveals extremely well the pole at origin, that appears already for the lower $N$, and is always associated with higher values for the absolute value of the
residua for all $N$s. On the other hand the pole at $p^2 = -0.5$ is not seen as clearly as the pole at the origin. However, 
for lower $N$ the dominant poles are located at the right $p^2 = - 0.5$ but, as $N$ is increased, it moves away from its right position and, for some $N$,
a zero of the approximant is associated with the pole position. We have checked that the identification of a single pole improves both
when the number of data points increases and when the statistical errors on the data become smaller. Further, exploring the distribution of poles and zeros
for the  complex $p^2$, see Fig. \ref{fig:Test_OffAxis_tree}, no stable positions are observed. The conclusion from studying the plots mentioned previously
is  that the analysis of the Pad\'e approximants data generated from Eq. (\ref{Eq:TestFuncs1}) suggests that a single pole should be associated with the data.
The sequences of Pad\'e approximants are able to reproduce the analytic structure of the original function.

\begin{figure}[t] 
   \centering
   \includegraphics[width=3in]{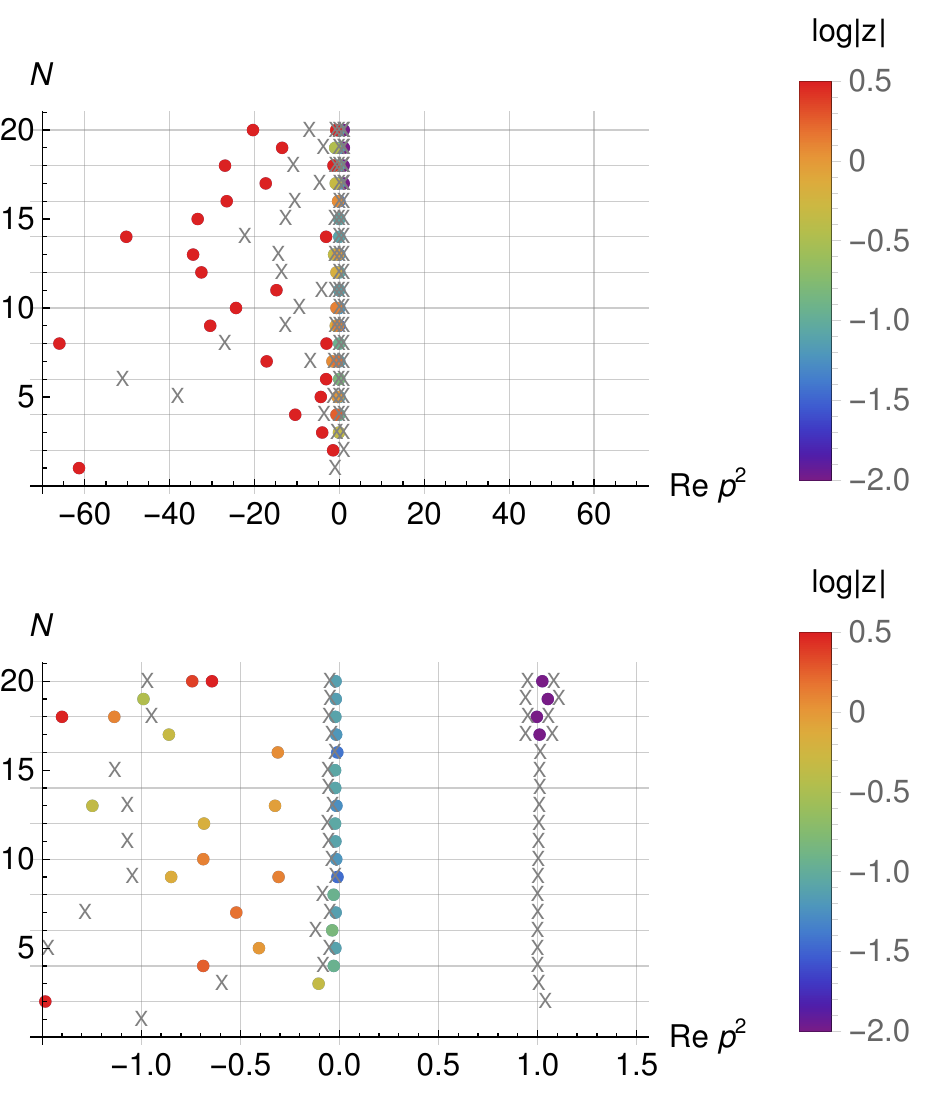} \\
   \includegraphics[width=3in]{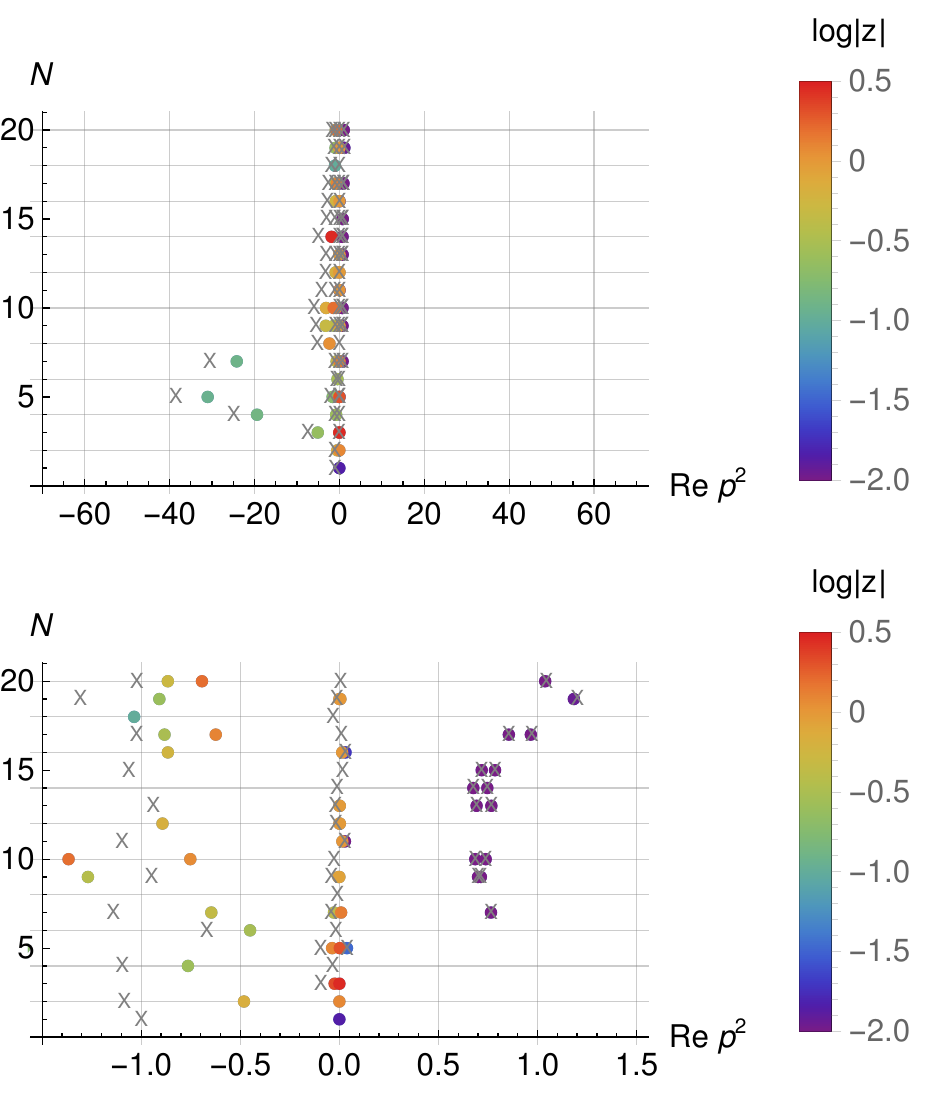}    
   \caption{The zeros and poles computed from the sequence of Pad\'e approximants for the data generated using Eq. (\ref{Eq:TestFuncs2}) with 
                 $m^2 = 0$ (top two plots) and  Eq. (\ref{Eq:TestFuncs3}) with $\omega = 0.3$ (bottom two plots).}
   \label{fig:Test_OnAxis_Log}
\end{figure}

The remaining functions (\ref{Eq:TestFuncs2}) and (\ref{Eq:TestFuncs3}) have branch cuts at on-axis negative values of $p^2$.
For these functions, the sequence of zeros and poles, along the real axis $p^2$, coming from the sequence of Pad\'e analysis can be seen
in Fig. \ref{fig:Test_OnAxis_Log}. 
In both cases there is a stable sequence of close poles and zeros that starts at the branch point $p^2 = 0$ and move towards the negative $p^2$ axis. 
If for the pure logarithm function, the poles with the largest residuum are not those close to the origin, for the perturbative like solution (\ref{Eq:TestFuncs3})
the position of the dominant pole is preferably at the true pole position.

Similarly as for the function (\ref{Eq:TestFuncs1}), one can look to the set of poles and residua as in Fig. \ref{fig:Test_OffAxis_tree} with the
results repeating the pattern observed in this Fig. These results suggest that, indeed, the function hidden in the data has no poles for complex $p^2$.
More, these results suggests that in a sequence of Pad\'e approximants a branch cut is identified by a sequence of zeros and poles with large residua.
Again, the Pad\'e analysis seems to be able to identify a branch cut and a single pole on top of a branch point.

\begin{figure}[t] 
   \centering
   \includegraphics[width=3in]{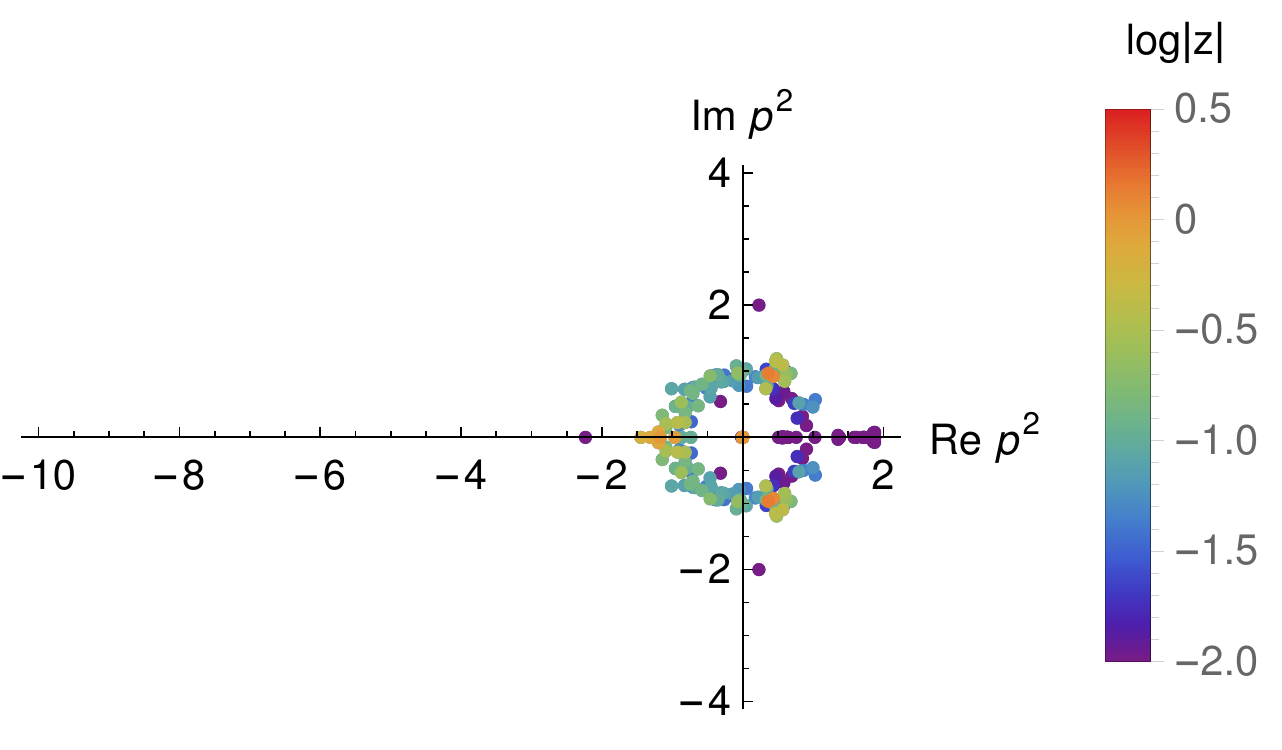} \\
   \includegraphics[width=3in]{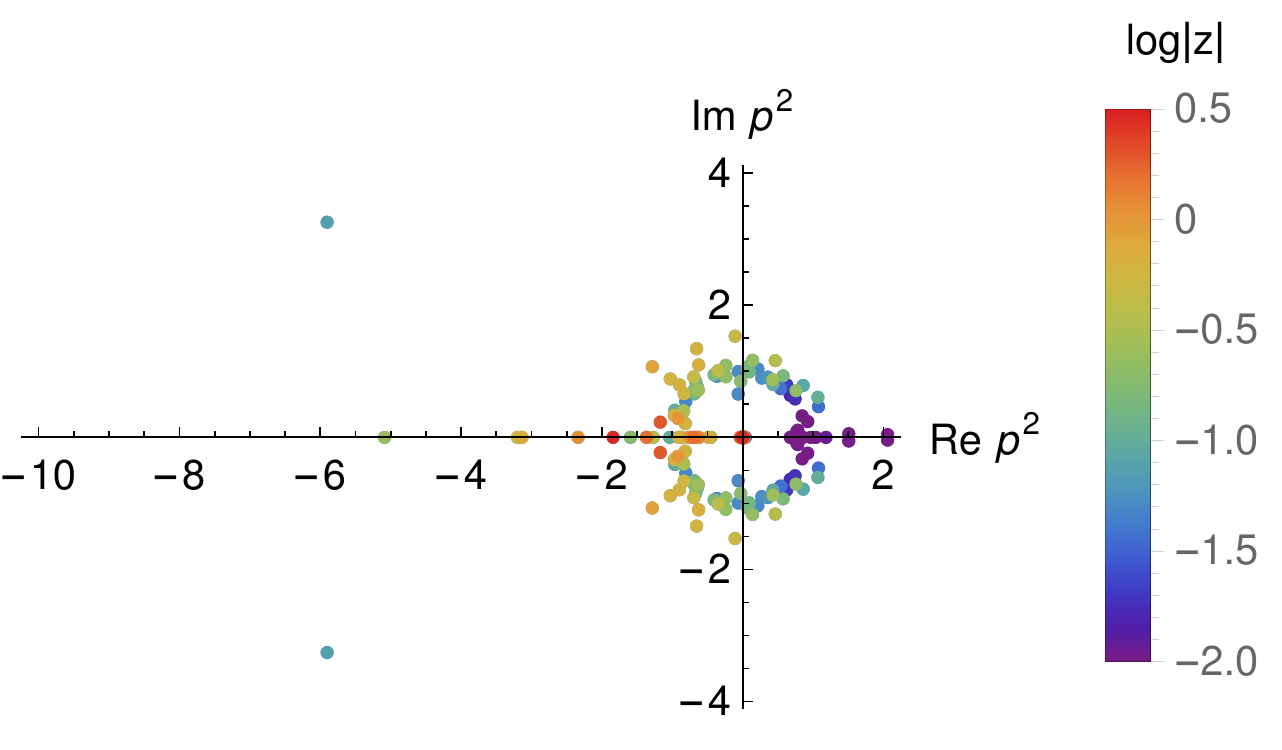}    
   \caption{Full set of poles  for all $N$, from the Padé analysis for the data generated using Eq. (\ref{Eq:TestFuncs1}) 
                  with $m^2 = 0$ (top) and  Eq. (\ref{Eq:TestFuncs3}) with $\omega = 0.3$ (bottom).}
   \label{fig:Test_All_p2_pert}
\end{figure}

For completeness in Fig. \ref{fig:Test_All_p2_pert} we report the full set of poles in the entire complex plane and for all $N$, 
as given by the Pad\'e approximant analysis for the data associated with (\ref{Eq:TestFuncs1}) (top plot) and with (\ref{Eq:TestFuncs3}) (bottom plot).
The full set of poles and zeros have a complex pattern and their absolute value of the residua have an hierarchy of  values (see the legend at the right side).
We call the readers attention to the position of the poles with the largest absolute value of the residua.
Fig. \ref{fig:Test_All_p2_pert} also show the Froissart doublets that necessarily appear at sufficiently large $N$. 

Besides the studies using the sets of data generated from Eqs. (\ref{Eq:TestFuncs1}) to (\ref{Eq:TestFuncs3}), we also investigated the outcome of
a standard Pad\'e analysis, i.e. on the results that uses a series expansion for the gluon propagator functional given by renormalisation group improved perturbation
theory
\begin{equation}
   D(p^2) = \frac{1}{p^2} \left( \omega \, \log \frac{p^2}{\Lambda^2_{QCD}} + 1 \right)^\gamma \ ,
   \label{Eq:Dp2_Analytic}
\end{equation}
where $\gamma = -13 / 22$ is the gluon anomalous dimension for pure Yang-Mills theory.  In order to perform the Pad\'e analysis of this function 
we took, for the various constants the values used in \cite{Dudal:2018cli} to describe the lattice data, 
namely $\omega = 33 \, \alpha_s / 12 \, \pi$, with $\alpha_s = 0.3837$ and $\Lambda_{QCD} = 0.425$ GeV.
A standard analysis show that the Pad\'e approximants reproduce the pole at $p^2 = 0$, that appears as a stable point at the right location, 
for both \pade{N}{N} and \pade{N-1}{N} sequences. 
Further, a structure of poles and zeros on the negative side of the real $p^2$ axis, that start at $p^2 = 0$, simulate the branch cut along the negative real axis 
similarly as in Fig. \ref{fig:Test_OnAxis_Log}.
Moreover, the analysis of the sequences \pade{N}{N}, \pade{N-1}{N} and \pade{N-2}{N} give a quite small coefficient associated with
the largest power in the denominator for \pade{N}{N} Pad\'e sequences, compared with the remaining coefficients, 
and the two sequences \pade{N-1}{N} and \pade{N-2}{N} result in essentially the same quality for the approximant. 
Our interpretation for this results being that the Pad\'e approximant suggests that, at large momentum, the gluon propagator behaves as a $1/(p^2)^\iota$ 
with $\iota$ being somewhere between one and two, i.e. the Pad\'e approximants are sensitive to the $\log$ corrections of the tree level perturbation theory.

For the standard Pad\'e analysis, we also considered the case where the simple pole at the origin was regularised by a mass term and where the $\log$ 
was also regularised by a constant mass term. In general, we found that the Pad\'e approximants, taken from the series expansions, are able to reproduce the 
appropriate analytic structures. However, if the mass term that regularises the $\log$ becomes complex valued the Pad\'e analysis was able to
identify correctly the branch point but do not predict correctly the position of the branch cut.

The study of the test functions show that the sequences of Pad\'e approximants can provide a reliable glimpse of the analytic structure of certain types of
functions. The analysis performed for $D_1(p^2)$, $D_2(p^2)$ and $D_3(p^2)$ 
will certainly guide us in the understanding of the analytic structure of the lattice propagator 
data using sequences of Pad\'e approximants.

\section{Pad\'e approximants and the lattice propagators \label{Sec:general}}

Let us try to understand what type of Pad\'e approximants should we use to describe the lattice gluon propagator. Although focusing now only on the gluon propagator, 
similar reasonings apply to the ghost propagator with minimal changes. The one-loop renormalisation group improved prediction for the gluon propagator
(Euclidean space) is given in Eq. (\ref{Eq:Dp2_Analytic}) 
where $\omega = 11 \, N \, \alpha_s(\mu^2) / 12 \, \pi$, $\alpha_s ( \mu^2)$ is the strong coupling constant defined at the renormalisation scale $\mu$
and $\gamma_{gl} = -13/22$ is the gluon anomalous dimension. 
This expression can be compared with gluon lattice data to check if the lattice data is sensitive to the logarithm correction to the tree level propagator.
Herein, in order to investigate for the presence of the log behaviour in the lattice data, we will consider the propagator computed with the 
ensembles of gauge configurations published in \cite{Dudal:2018cli}.
The lattice data is renormalised in the MOM-scheme through the condition
\begin{equation}
    \left. D(\mu^2) \right|_{\mu = 3 ~ GeV} = \frac{1}{\mu^2} \ .
    \label{Eq:renormalisacao}
\end{equation}
Details of the simulation and of  the lattice setup can be found in \cite{Dudal:2018cli}. 

\begin{figure}[t]
   \centering
   \includegraphics[scale=0.31]{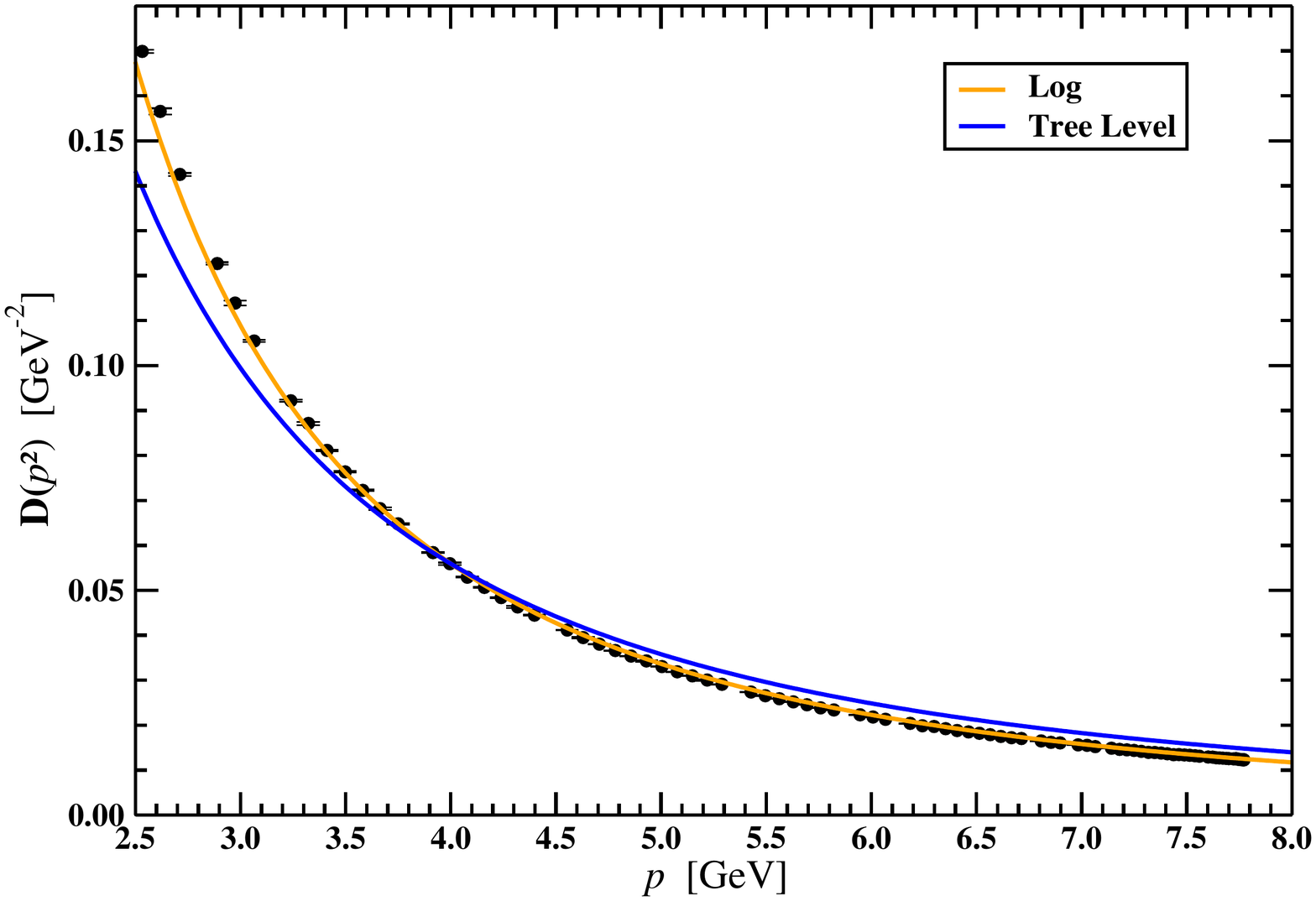}  
   \includegraphics[scale=0.31]{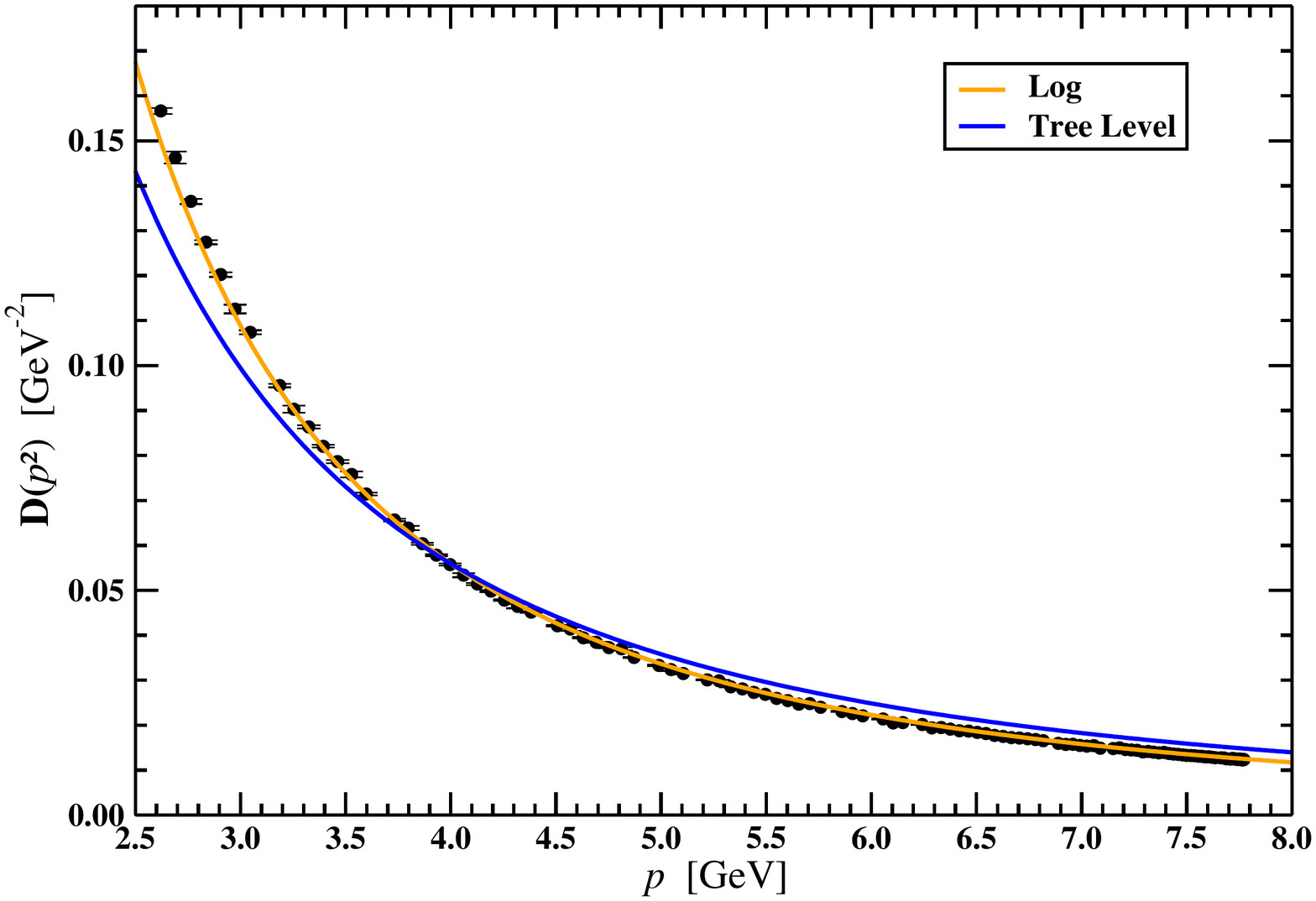}  
   \caption{The ``high'' momenta lattice Landau gauge gluon propagator for a simulation on a $64^4$ lattice (top) and on a $80^4$ lattice (bottom) compared
                 to the tree level expression $1/p^2$ and the one-loop renormalization group improved prediction as given in (\ref{Eq:Dp2_Analytic}). 
                 The lattice data shown is renormalised in the MOM-scheme at $\mu = 3$ GeV. The tree level and the one-loop expressions were matched to
                 the lattice data for $\mu = 4$ GeV. The one-loop expression for the gluon propagator was computed with $\Lambda_{QCD} = 0.425$ GeV,
                 $\alpha_s( 3 \mbox{ GeV} )$ $= 0.3837$ as in \cite{Dudal:2018cli}, where the details about the lattice simulation can be found. }
   \label{fig:gluon_high_tree_1loop}
\end{figure}

In Fig. \ref{fig:gluon_high_tree_1loop} we compare the renormalised gluon propagator with both the tree level expression
\begin{equation}
  D(p^2) = \frac{1}{p^2}
  \label{Eq:gluon_tree0}
\end{equation}
and Eq. (\ref{Eq:Dp2_Analytic}). The overall scale for the expressions (\ref{Eq:gluon_tree0}) and (\ref{Eq:Dp2_Analytic}) is fixed by demanding
that the functional forms match the lattice data at $p = 4$ GeV; note that the matching is not performed exactly at the renormalisation
scale. The numerical values of the various parameters used to build the curves are reported in the caption of Fig. \ref{fig:gluon_high_tree_1loop}.

The curves in Fig. \ref{fig:gluon_high_tree_1loop}  show that (\ref{Eq:Dp2_Analytic}) is on top of the lattice data for momenta $p \sim 3$ GeV
and above, while the tree level expression (\ref{Eq:gluon_tree0}) shows clear deviations from the lattice data for the range of momenta considered. 
We take this as an indication that the high precision lattice data of \cite{Dudal:2018cli} identifies correctly the one-loop logarithmic correction given in Eq.
(\ref{Eq:Dp2_Analytic}). 

For our purpose, i.e. the investigation of the analytic structure of the propagators, the lattice simulations will provide a set of $D(p^2)$ for real Euclidean 
$p^2$ that will be approximated by ratios of polynomials. In particular we will consider the Pad\'e approximants \pade{M}{N}
\begin{equation}
   D(p^2) \approx \frac{Q_M(p^2)}{R_N(p^2)} \ .
   \label{Eq:usual_Pade_M_N}
\end{equation}
already mentioned previously in Eq. (\ref{Eq:usual_Pade_M_N_0}).
The per\-tur\-ba\-ti\-ve propagator shows a branch cut along the negative part of the real Euclidean $p^2$ axis and, to accomodate 
for such possibility, besides (\ref{Eq:usual_Pade_M_N}) it would be natural to look at approximants of the type \padelog{M}{N}{O}{S} given by
\begin{equation}
   D(p^2) \approx \frac{Q_M(p^2)}{R_N(p^2)} \, \left[ \omega \, \ln \frac{L_O(p^2)}{K_S(p^2)} + 1 \right]^{\gamma_{gl}} \ .
   \label{Eq:cut_Pade_M_N}
\end{equation}
In Eqs. (\ref{Eq:usual_Pade_M_N}) and (\ref{Eq:cut_Pade_M_N}) the polynomials $Q_M(p^2)$, $R_N(p^2)$, $L_O(p^2)$ and $K_S(p^2)$ are
defined as
\begin{eqnarray}
     Q_M(p^2) & = & q_0 + \cdots + q_M \, \left( p^2 \right)^M \ , \\
     L_N(p^2) & = & l_0 + \cdots + l_N \, \left( p^2 \right)^N \ , \\
     R_O(p^2) & = & 1 + \cdots + r_O \, \left( p^2 \right)^O \ , \\
     K_S(p^2) & = & 1 + \cdots + k_S \, \left( p^2 \right)^S \ .
\end{eqnarray}    
However, in practice, maybe due to the poor sensitivity to the variations of the coefficients that define the polynomials that appear in the logarithmic correction, 
it turns out that the minimisation of the $\chi^2$ using expression (\ref{Eq:cut_Pade_M_N}) is rather difficult to perform as the analytic structure
changes significantly as  $N$ is increased.
For these reasons we will omit the outcome of the analysis based on the use of Eq. (\ref{Eq:cut_Pade_M_N}).

The applications based on Pad\'e approximants use typically the diagonal and/or the near diagonal approximants. We follow the same rule
and, for the class of approximants given by Eq. (\ref{Eq:usual_Pade_M_N}), we will investigate the ratios of polynomials that have
$M = N$ and $M=N-1$. 
The motivation to set $M = N-1$ and not $M = N+1$ comes from results of perturbation theory, a behaviour that the approximant should reproduce at 
large $p^2$.
As for the test functions considered previously, for each Pad\'e approximant, the coefficients of the polynomials are computed looking at the 
(candidate) absolute minima for the $\chi^2$ defined  in Eq. (\ref{Eq:objectivefunction}).

\section{Pad\'e Approximants and the lattice Landau gauge gluon propagator \label{Sec:gluon}}

\begin{figure}[t]
   \centering
   \includegraphics[scale=0.3]{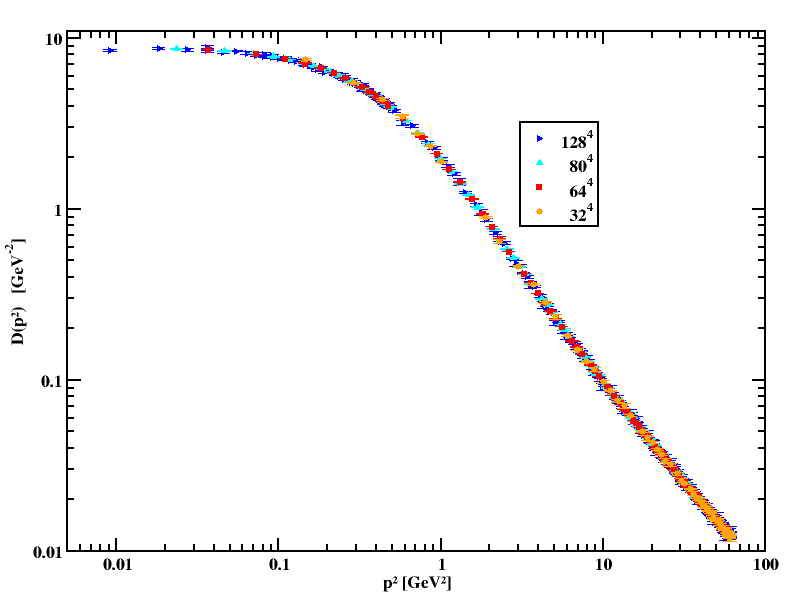} 
   \caption{Landau gauge gluon propagator used in the Pad\'e analysis.}
   \label{fig:gluon_data_all}
\end{figure}

\begin{figure*}[t]
   \centering
   \includegraphics[scale=0.8]{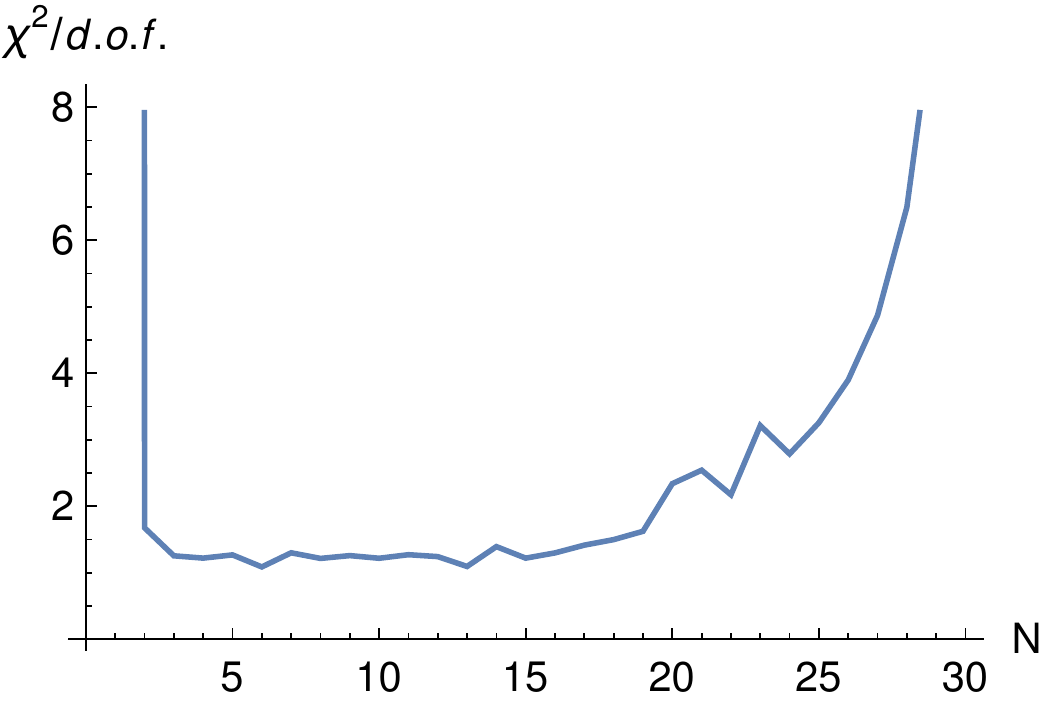}  ~~
   \includegraphics[scale=0.8]{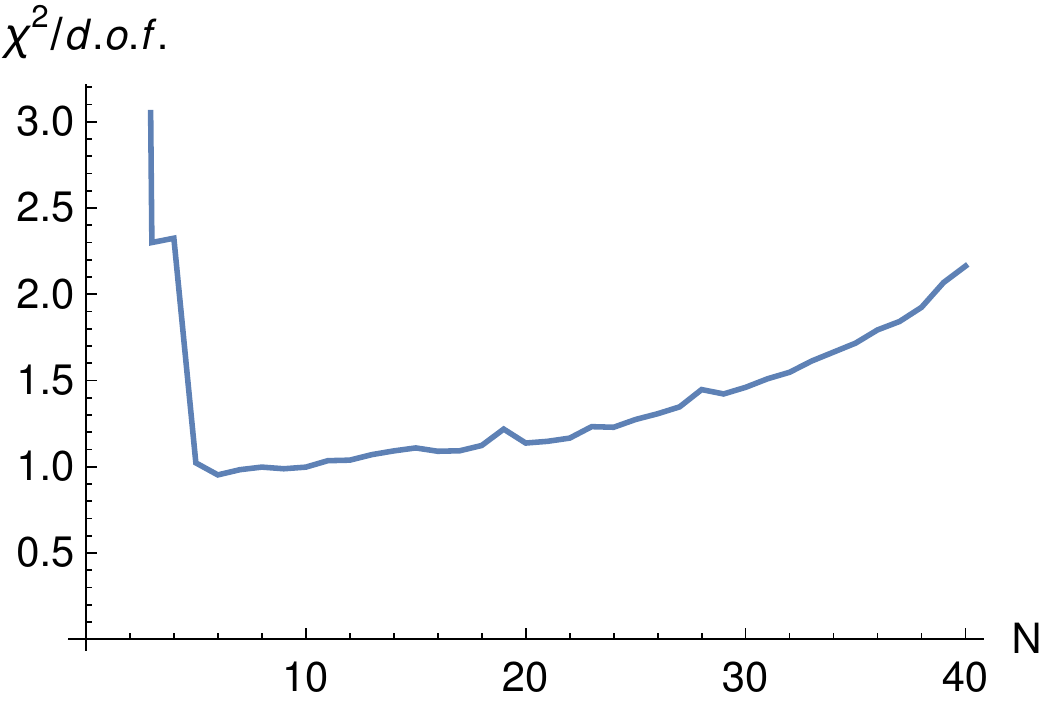}  \\ 
   \includegraphics[scale=0.8]{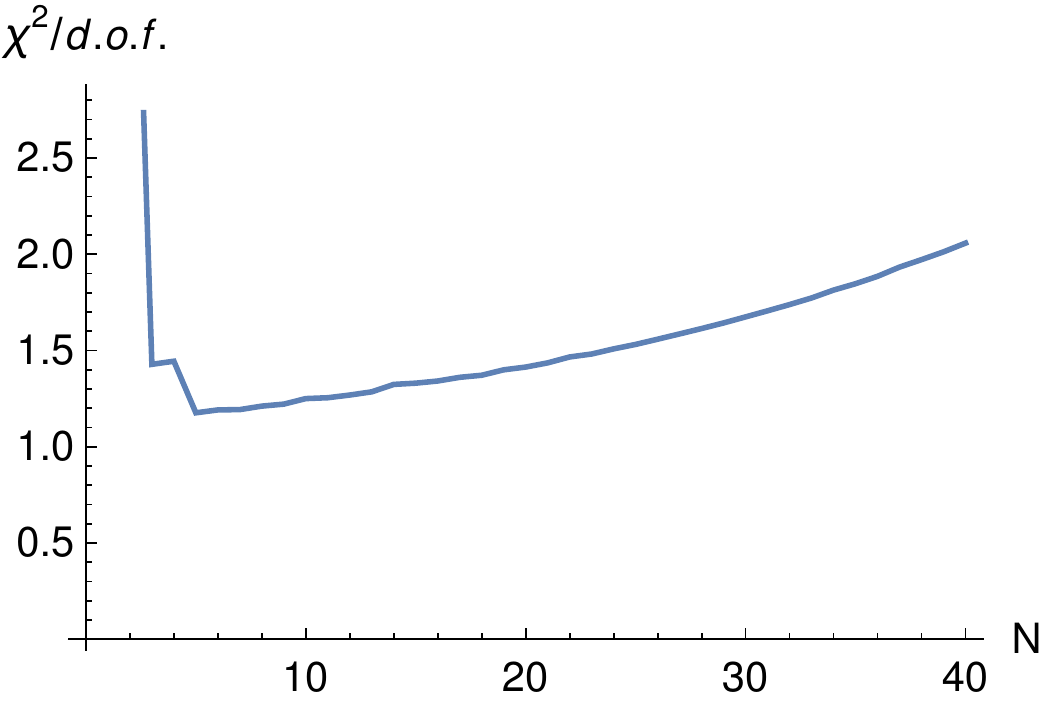}  ~~
   \includegraphics[scale=0.8]{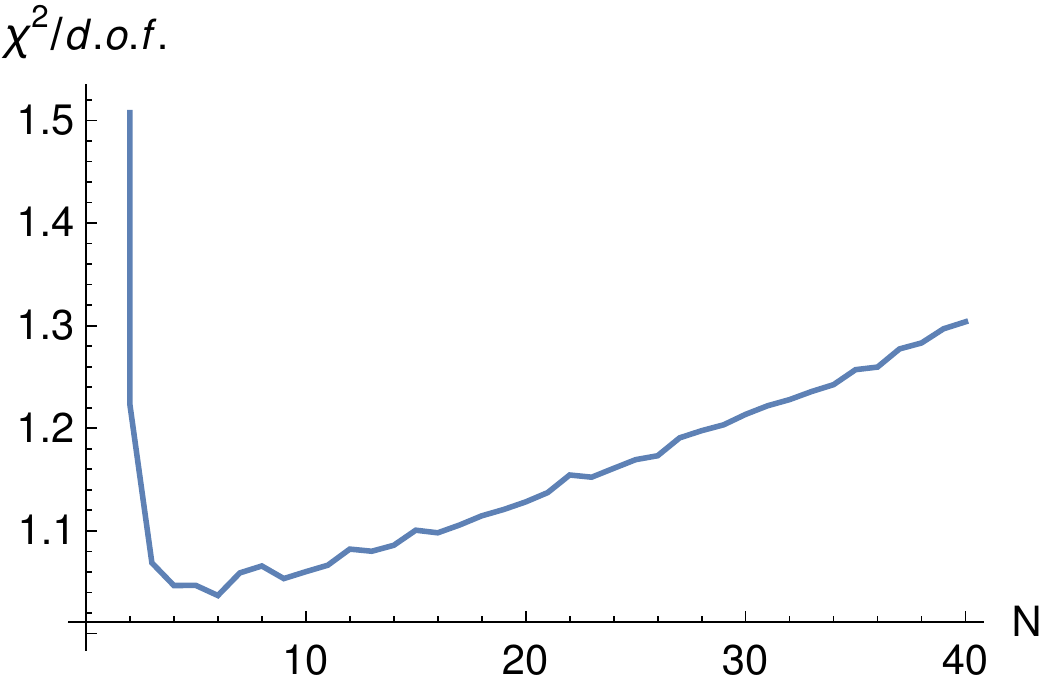}  
   \caption{The $\chi^2/d.o.f.$ as obtained in the minimisation process with the DE method. From top left to right bottom the the plots
                 refer to the minimisation of the $32^4$, of the $64^4$, of the $80^4$ and of the $128^4$ lattice data as a function of $N$ for
                 Pad\'e approximants of type \pade{N-1}{N}. The corresponding curves computed with the SA method are similar.}
   \label{fig:gluon_chi2_all}
\end{figure*}

For the investigation of the analytic structure of the Landau gauge lattice gluon propagator we rely on simulations performed
on hypercubic spacetime lattices using the Wilson gauge action for $\beta = 6.0$ at several physical volumes. The lattice data
considered is associated with simulations that use
(i) a $32^4$ lattice with 50 gauge configurations, published in \cite{Bicudo:2015rma}; 
(ii) a $64^4$ lattice with 2000 gauge configurations, published in \cite{Dudal:2018cli}; 
(iii) a $80^4$ lattice with 550 gauge configurations, published in \cite{Dudal:2018cli} and 
(iv) a $128^4$ lattice using 35 gauge configurations, published in \cite{Duarte:2016iko}. The physical volumes for the lattices are,
respectively, (3.25 fm)$^4$, (6.50 fm)$^4$, (8.13 fm)$^4$ and (13.01 fm)$^4$ for a lattice spacing of $a = 0.1016(25)$ fm.
The rationale to use the data from all these simulations being that it allows to have a better sensitivity to different regions of momenta and, in this
way, to be able to identify clearly possible structures in the complex plan. Indeed, the data from the $32^4$ simulation the major number of data points
has a $p \gtrsim 1$ GeV and by increasing the number of lattice points the number of infrared momenta is increased. 
All the lattice data reported here was renormalised in the MOM-scheme according to (\ref{Eq:renormalisacao}).

The renormalised gluon propagator data used in the Pad\'e analysis can be seen in Fig. \ref{fig:gluon_data_all}. All data sets are essentially compatible with each
other at one standard deviation level and, in this sense, they define a unique curve.
The exception being the zero momentum propagator, not seen in Fig. \ref{fig:gluon_data_all},
for the simulation performed on the smallest physical volume that is larger than the corresponding values for all the other simulations.
In order to the check for the finite volume effects and the level of statistical precision achieved by the various simulations we report the values of $D(0)$ for
all the data sets that is 10.64(38) GeV$^2$ for the $32^4$, 8.900(49) GeV$^2$ for the $64^4$, 8.847(99) GeV$^2$ for the $80^4$ and 8.98(39) GeV$^2$ for the 
$128^4$ simulation. The reader should note that, due to the way the propagator is computed, the zero momentum propagator has, typically, the largest statistical error
and, therefore, its contribution to the $\chi^2$ is smaller than the remaining momenta.

In Fig. \ref{fig:gluon_chi2_all} we report on the values of the $\chi^2/d.o.f.$ obtained when one uses the differential evolution method to minimise the $\chi^2$
for Pad\'e approximants of type \pade{N-1}{N}, as a function of the degree of the polynomial in the denominator. Although not shown, the corresponding
curves computed with the simulated annealing method are essentially indistinguishable. The data in Fig. \ref{fig:gluon_chi2_all}
reveal that by increasing the lattice size the value of $\chi^2/d.o.f.$ decreases. In all cases, the minimisation results in acceptable values for the reduced
$\chi^2$. The exception are the outcome of the minimisations for the smallest lattice when $N \gtrsim 20$ that have large $\chi^2/d.o.f.$

\begin{figure*}[t]
  \centering 
   \subfigure[$32^4$]{\includegraphics[scale=0.8]{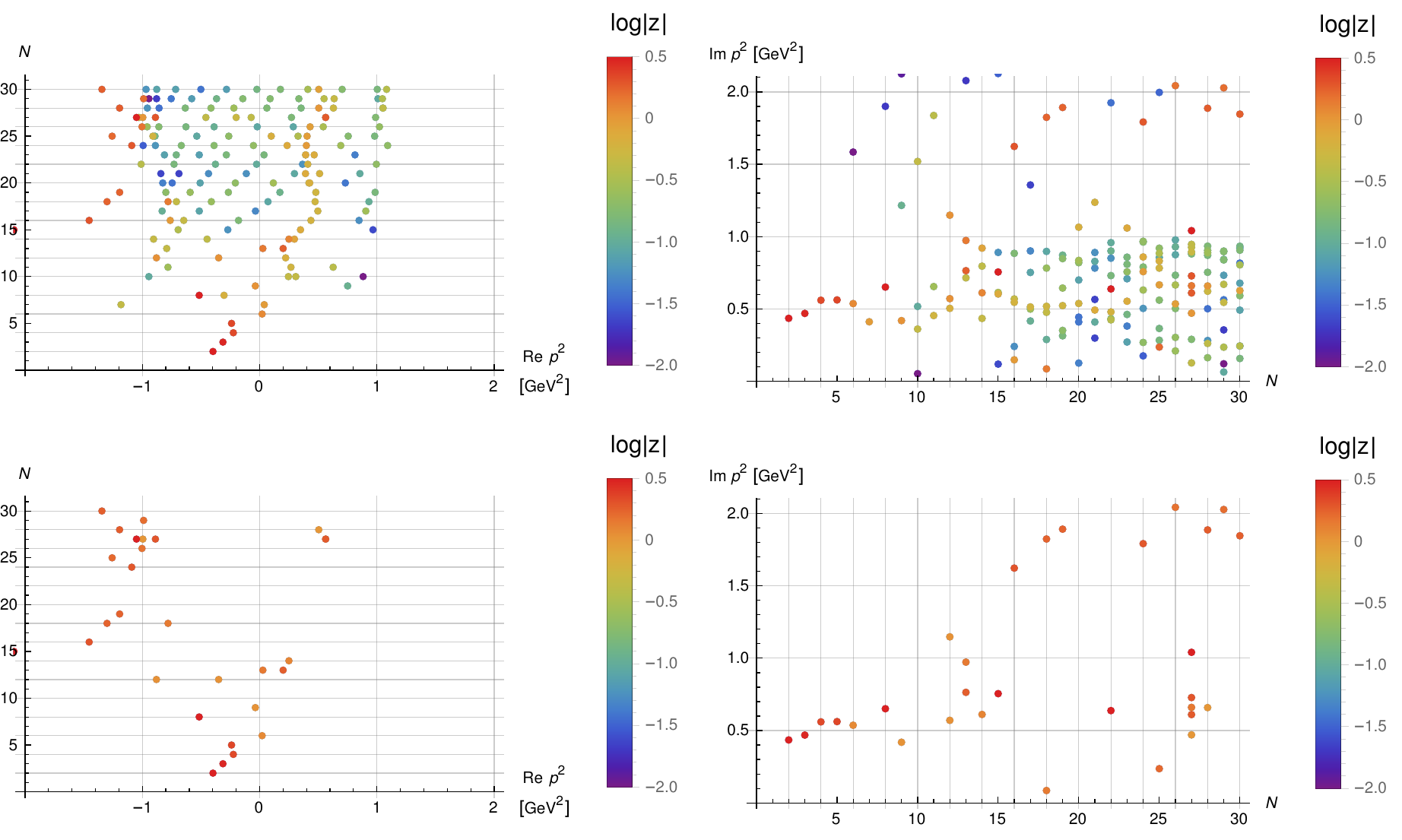}} \\
   \subfigure[$64^4$]{\includegraphics[scale=0.8]{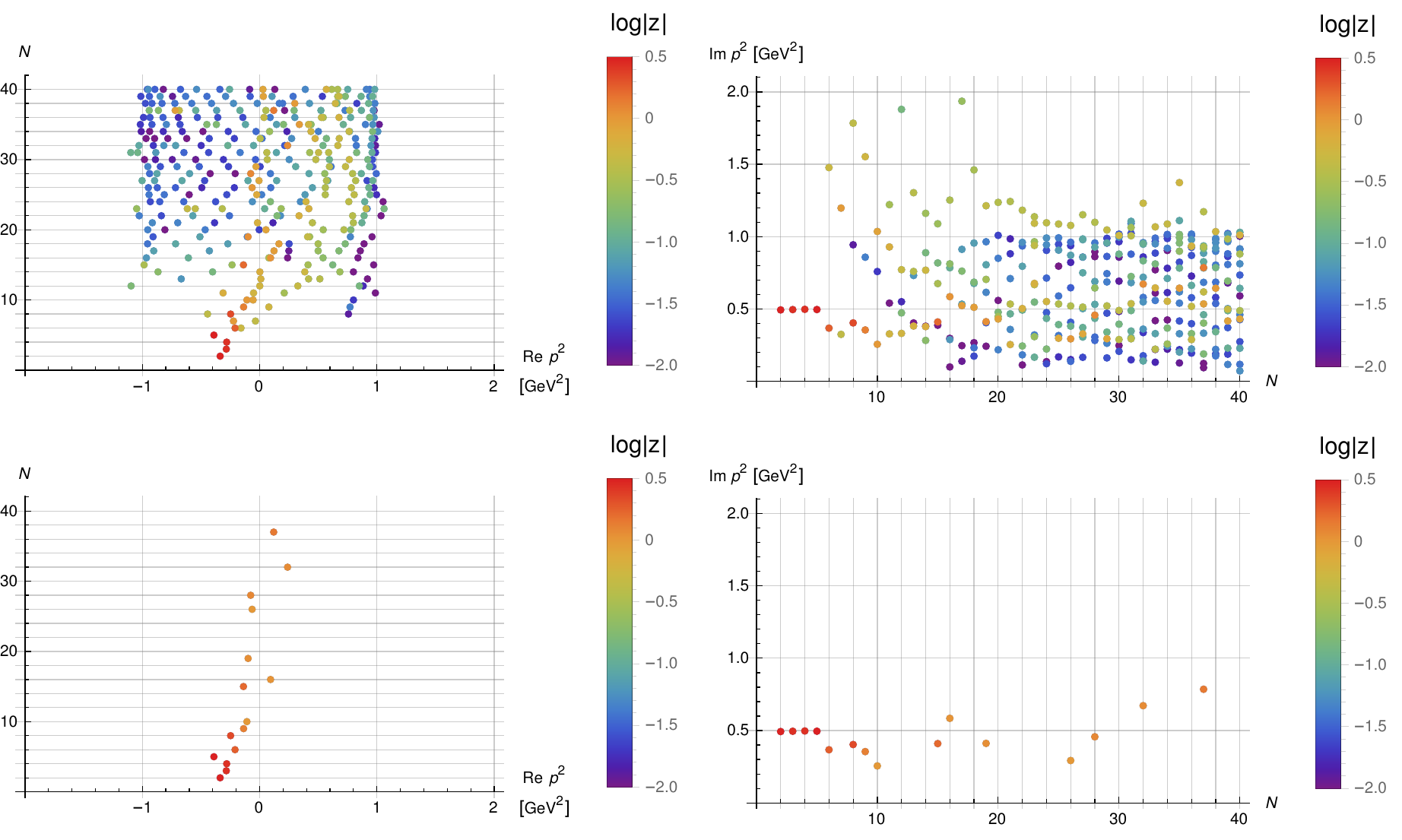}} 
   \caption{Evolution of the poles for complex momenta given by the Pad\'e approximants \pade{N-1}{N} and computed with the
                 differential evolution minimisation method. The scale on each plot refers to the absolute value of the residua for each pole.}
   \label{fig:gluon_DE_complex_1}
\end{figure*}
   
\begin{figure*}[t]
  \centering    \subfigure[$80^4$]{\includegraphics[scale=0.8]{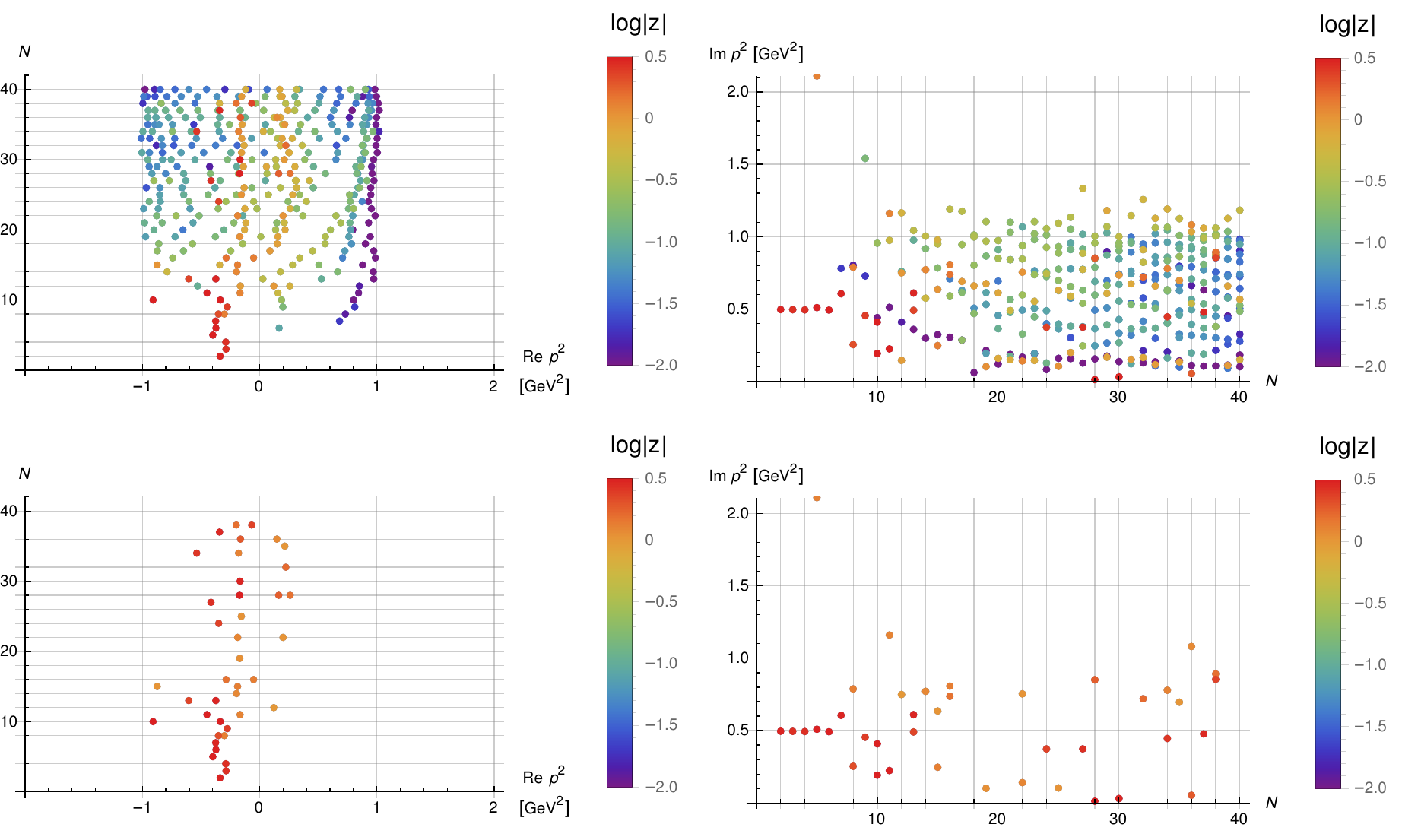}} ~
   \subfigure[$128^4$]{\includegraphics[scale=0.8]{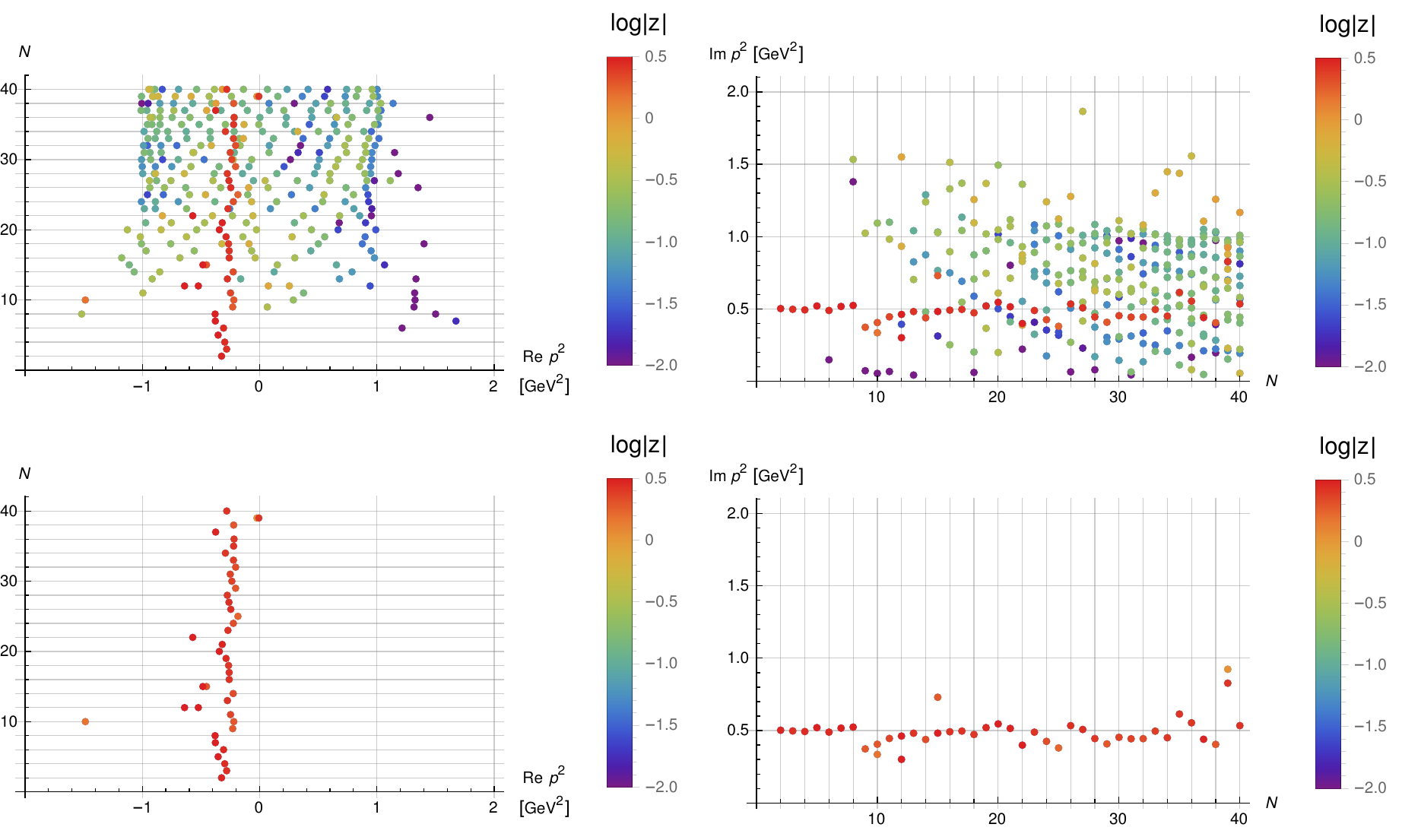}}
   \caption{The same in Fig. \ref{fig:gluon_DE_complex_1} for the two largest lattices.}
   \label{fig:gluon_DE_complex_2}
\end{figure*}

\begin{figure*}[t]
  \centering 
   \subfigure[$32^4$]{\includegraphics[scale=0.8]{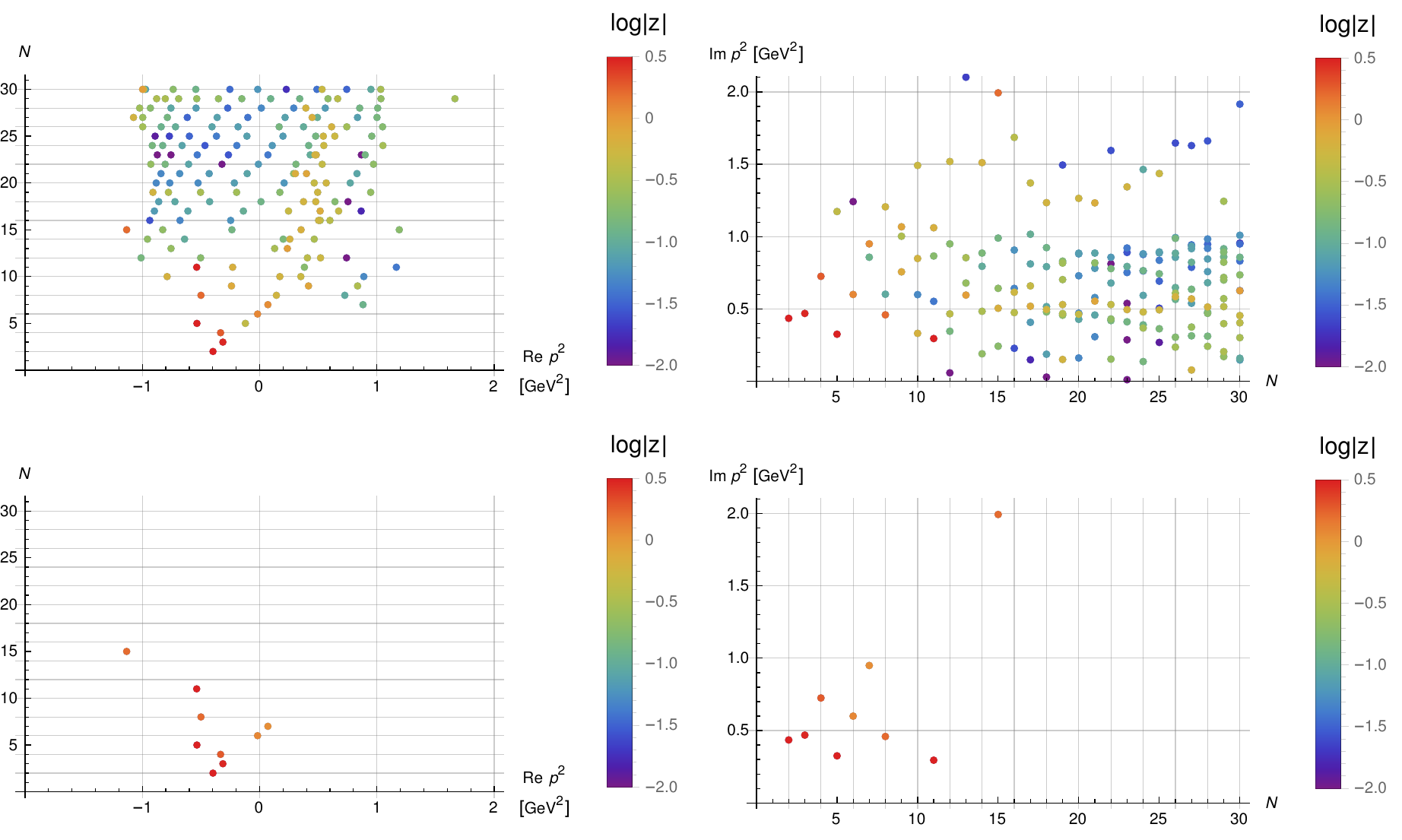}} ~
   \subfigure[$64^4$]{\includegraphics[scale=0.8]{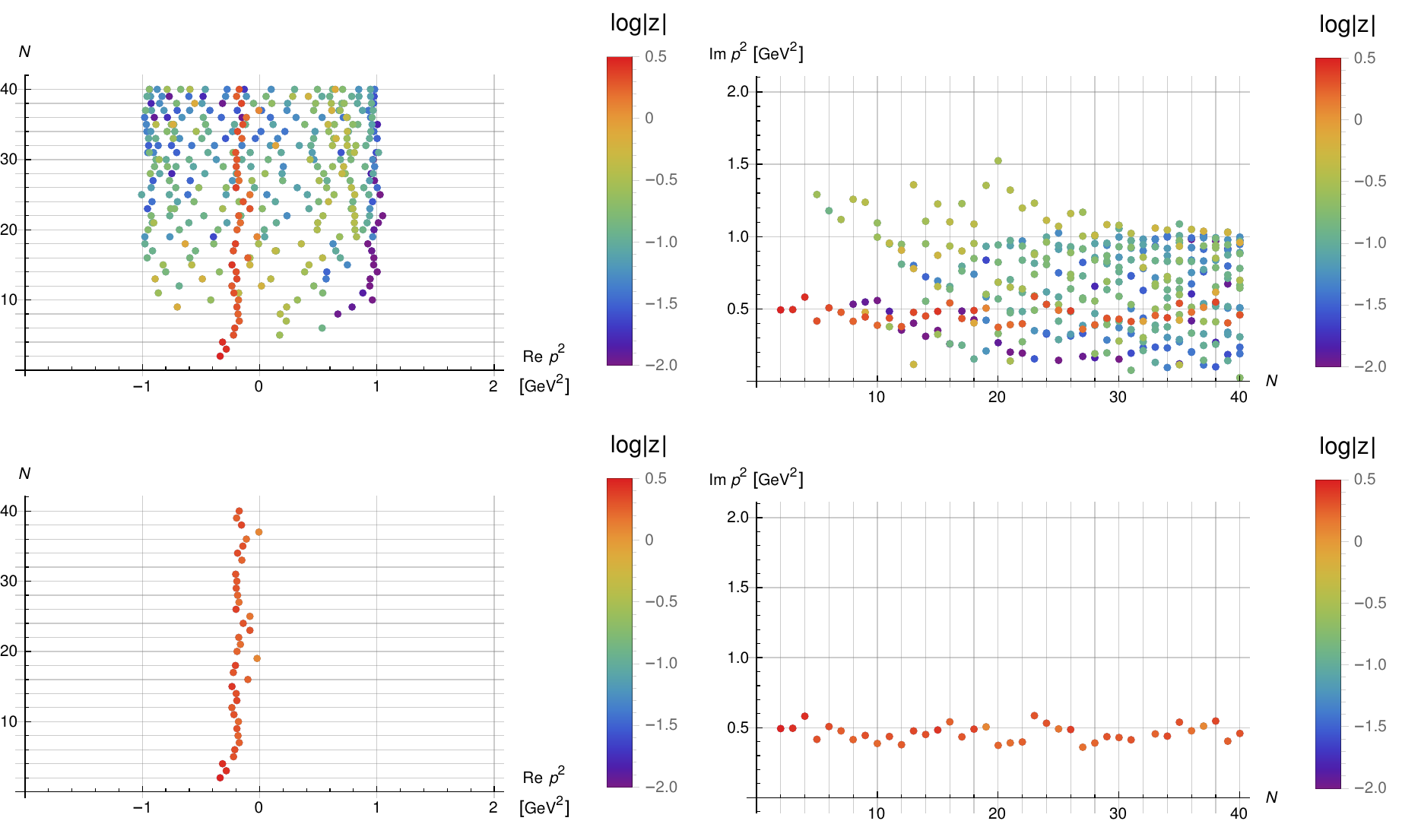}} \\
   \caption{Evolution of the poles for complex momenta given by the Pad\'e approximants \pade{N-1}{N} and computed with the
                 simulated annealing minimisation method. The scale on each plot refers to the absolute value of the residua for each pole.}
   \label{fig:gluon_SA_complex_1}
\end{figure*}

\begin{figure*}[t]
  \centering 
   \subfigure[$80^4$]{\includegraphics[scale=0.8]{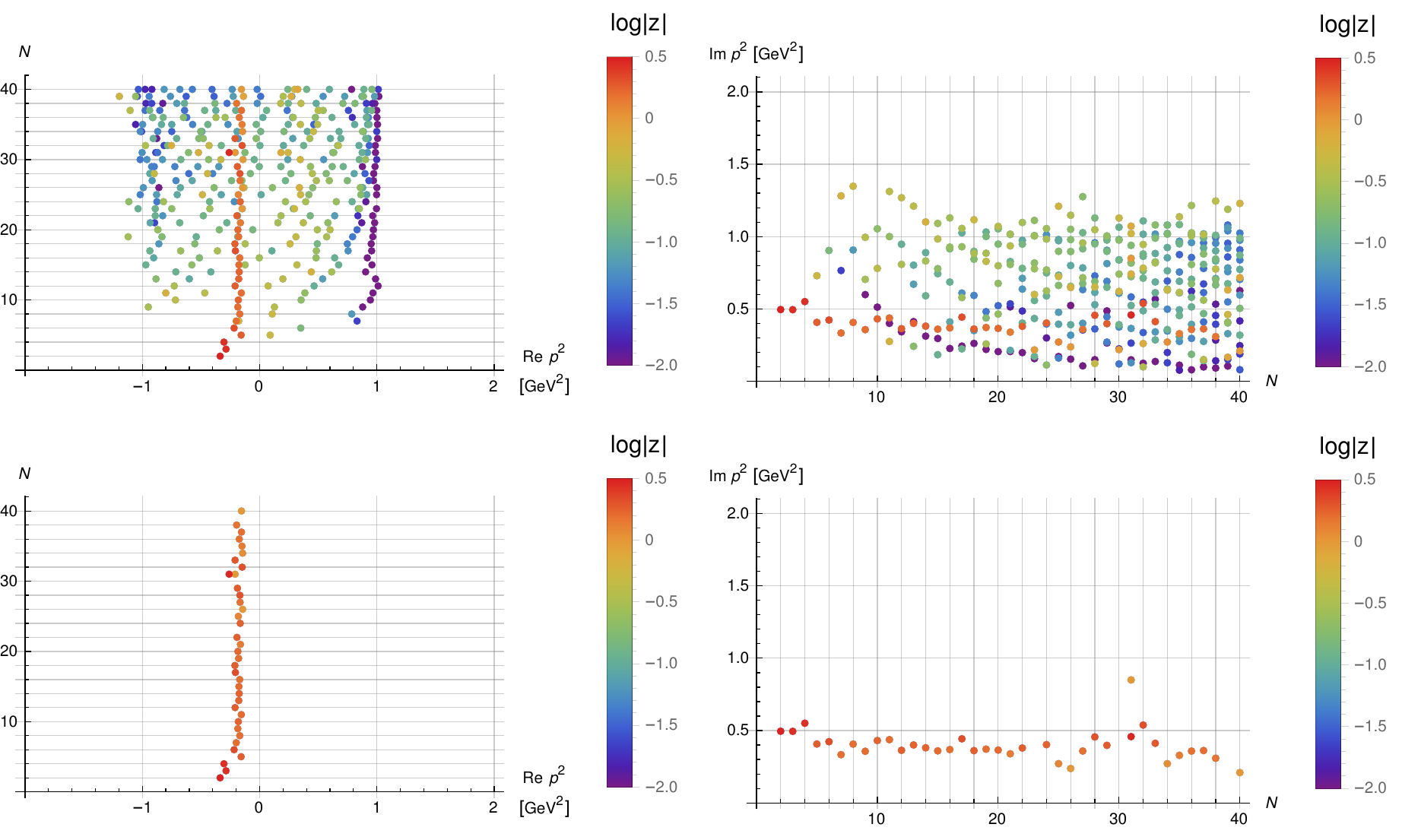}} ~
   \subfigure[$128^4$]{\includegraphics[scale=0.8]{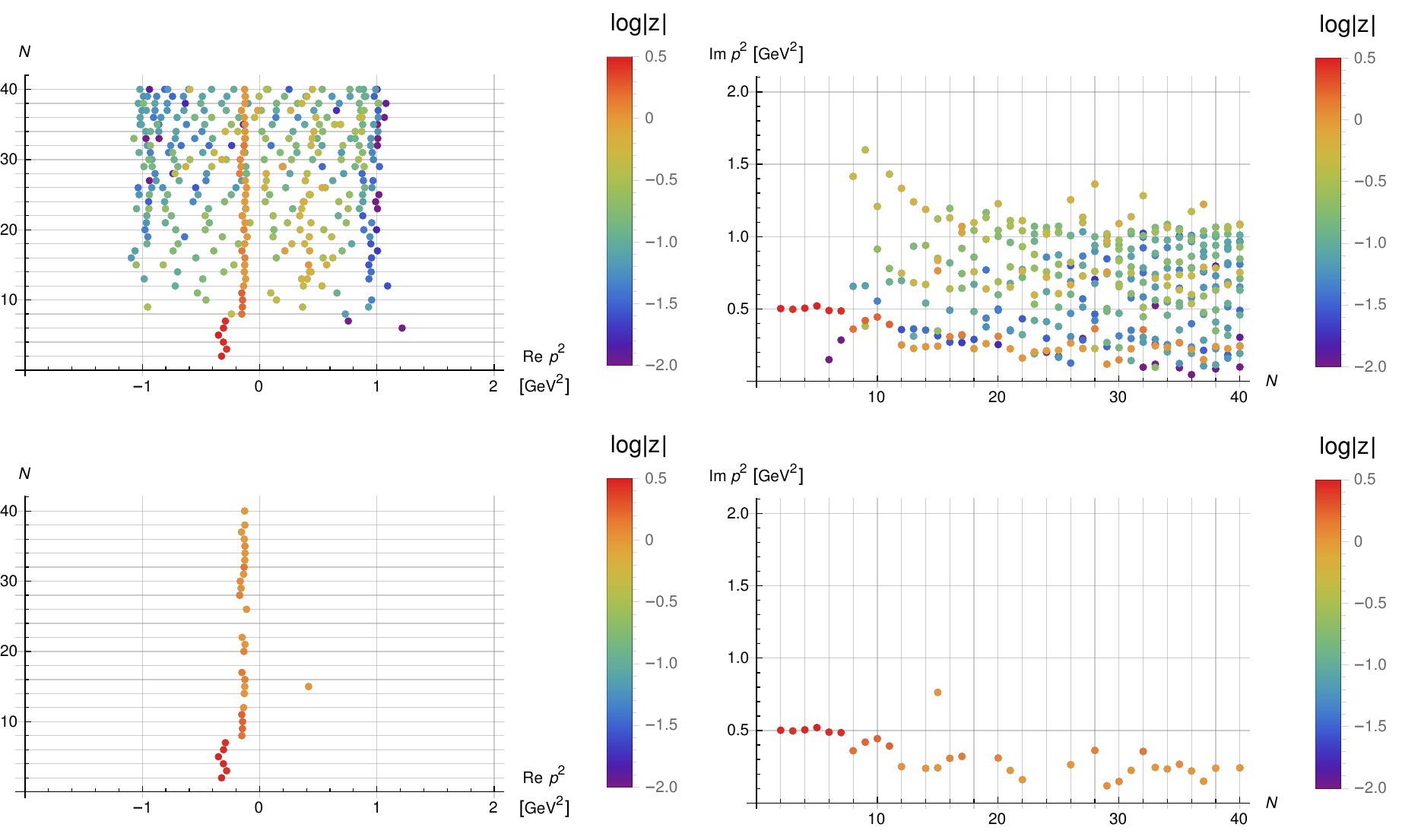}}
   \caption{The same in Fig. \ref{fig:gluon_SA_complex_1} for the two largest lattices.}
   \label{fig:gluon_SA_complex_2}
\end{figure*}

In Figs. \ref{fig:gluon_DE_complex_1} to \ref{fig:gluon_SA_complex_2} the poles of the propagators for complex momenta as 
given by the Pad\'e approximants \pade{N-1}{N}  are reported. Figs. \ref{fig:gluon_DE_complex_1} and \ref{fig:gluon_DE_complex_2} summarise
the results computed with the differential evolution method, while Figs. \ref{fig:gluon_SA_complex_1} and \ref{fig:gluon_SA_complex_2} shown
the outcome of the minimisation when using the simulated annealing method.
The scales in the r.h.s. of the Figs. refers to the absolute value of the residua of each pole. In all cases, 
the dominant poles, i.e those with the highest absolute value for their residua, are associated with the color red and
those poles with the smaller residua appear in dark blue.
For each lattice size all the Figs. have two sets of plots. The upper plot reports all the poles for complex $p^2$ as given by the Pad\'e approximants.
In the lower ones only the poles with the higher residua are shown, i.e. it includes the poles whose absolute value for the residuum is such that $\log |Z| > 0$.
In the Pad\'e sequences the poles appear always as pairs of complex conjugate $p^2$ values with the same $|Z|$.
The Figs. only show the poles that have $\Im( p^2 ) > 0$.

\begin{figure*}[t]
  \centering 
   \subfigure[$32^4$ - DE]{\includegraphics[scale=0.8]{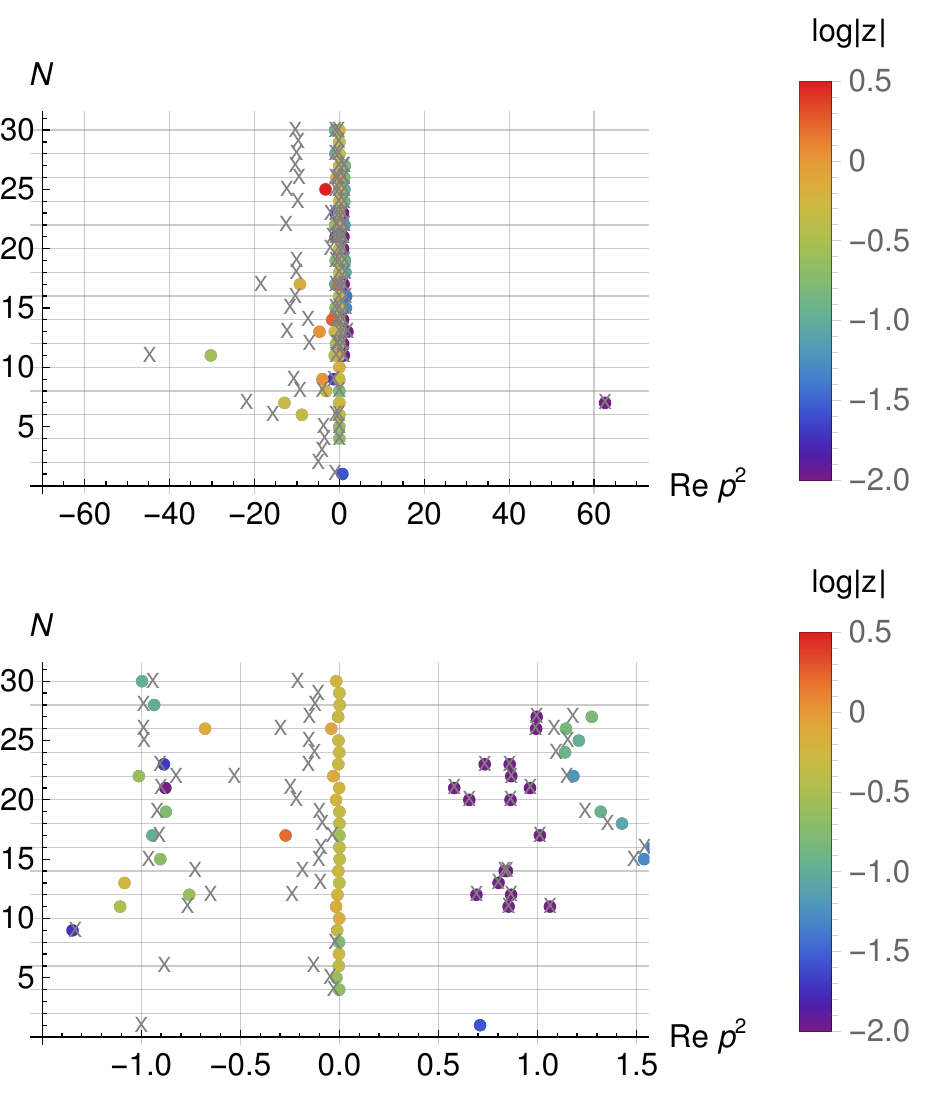}} ~
   \subfigure[$32^4$ - SA]{\includegraphics[scale=0.8]{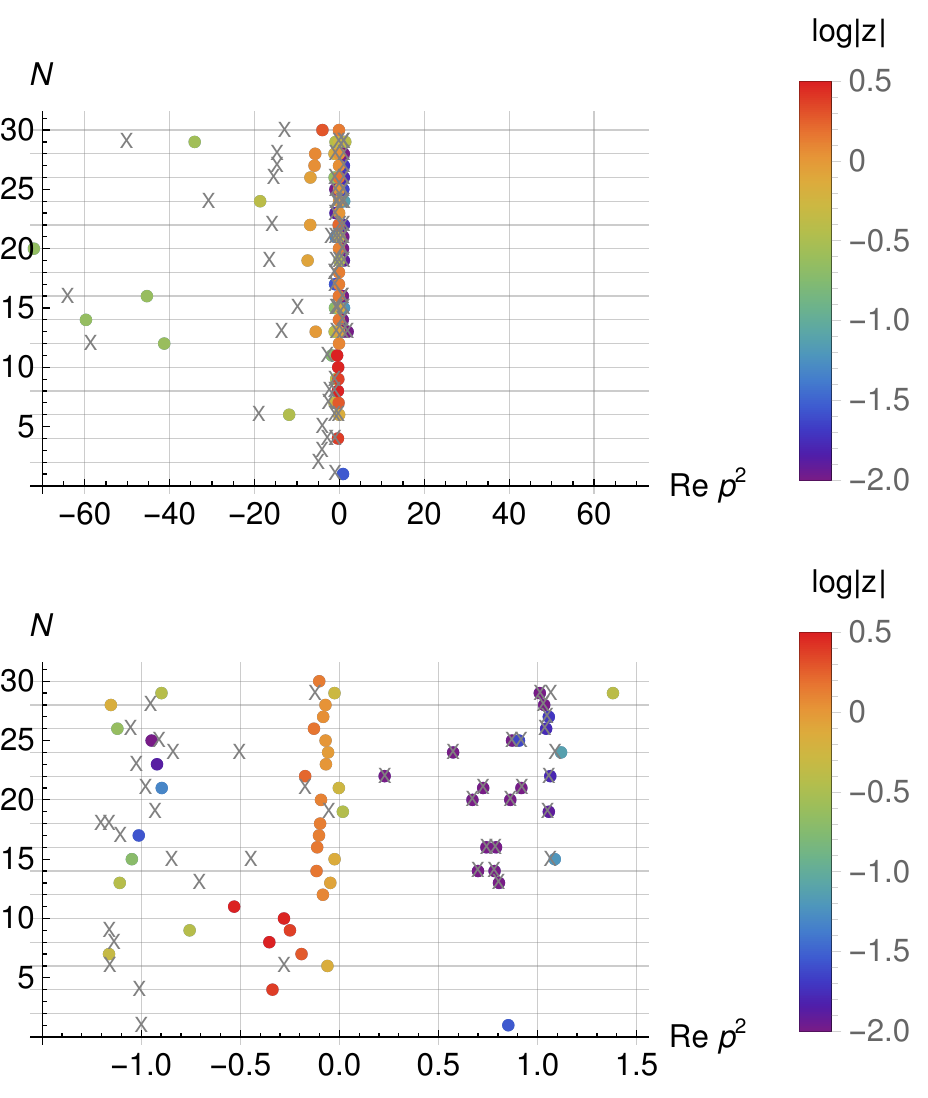}} \\
   \subfigure[$64^4$ - DE]{\includegraphics[scale=0.8]{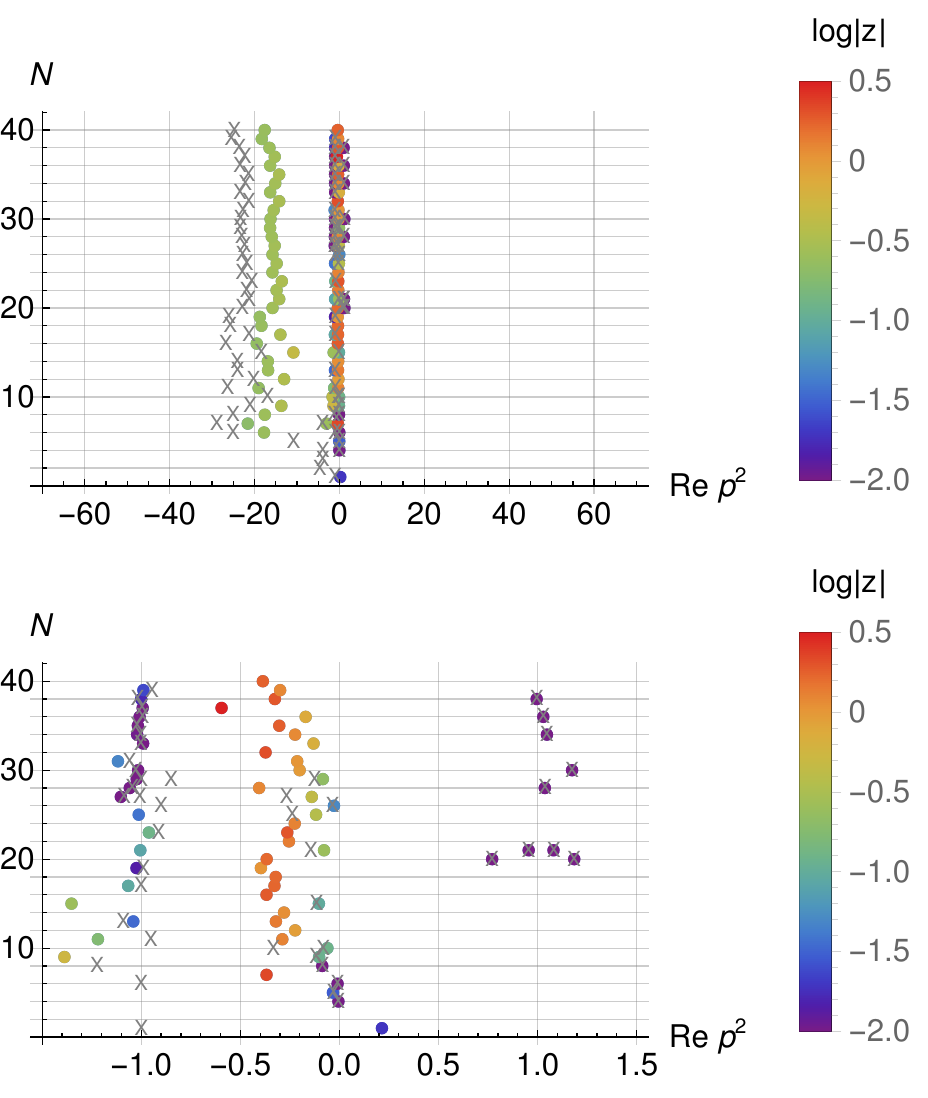}} ~
   \subfigure[$64^4$ - SA]{\includegraphics[scale=0.8]{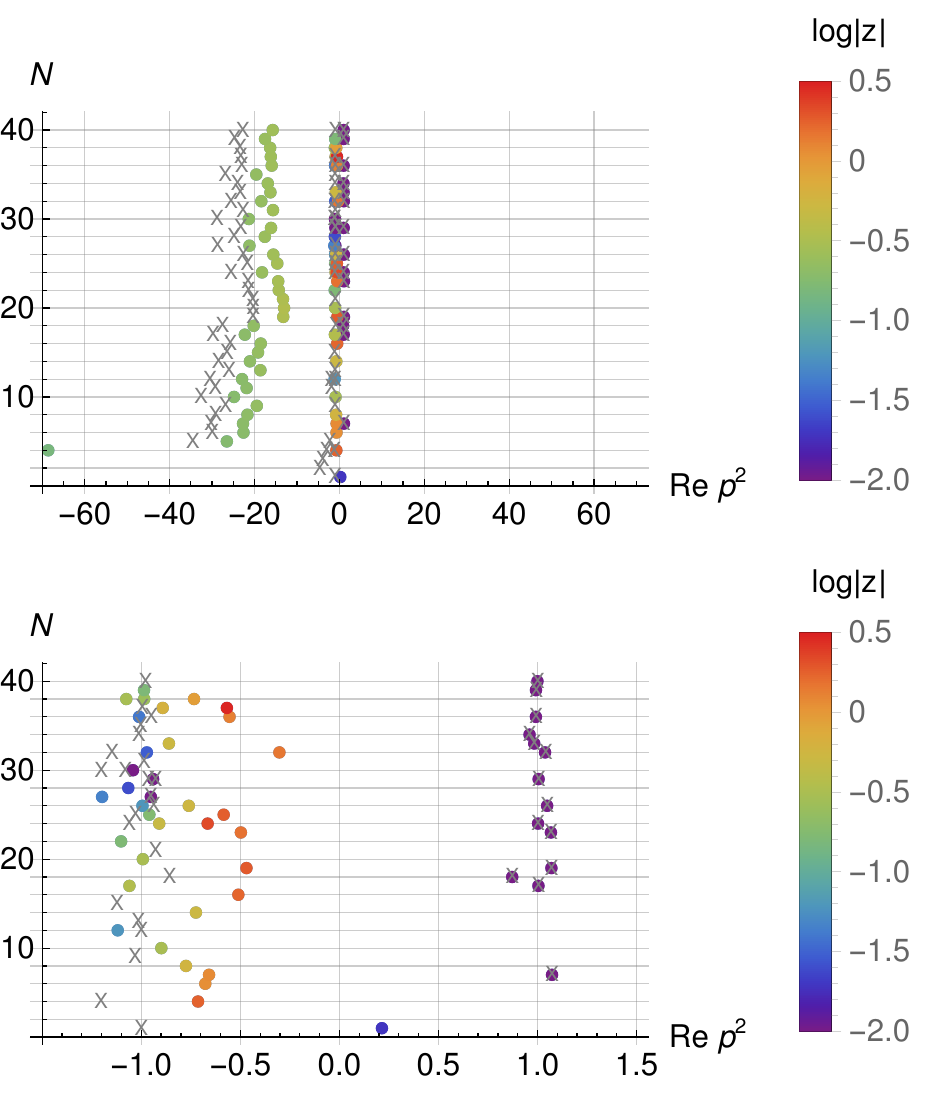}}
   \caption{Poles (circles) and zeros (crosses) from the Pad\'e approximants \pade{N-1}{N} at the real $p^2$ axis, computed using the
                 differential evolution and simulated annealing minimisation methods, for the two smallest lattices.}
   \label{fig:gluon_DE_onaxis_small}
\end{figure*}

\begin{figure*}[t]
  \centering 
   \subfigure[$80^4$ - DE]{\includegraphics[scale=0.8]{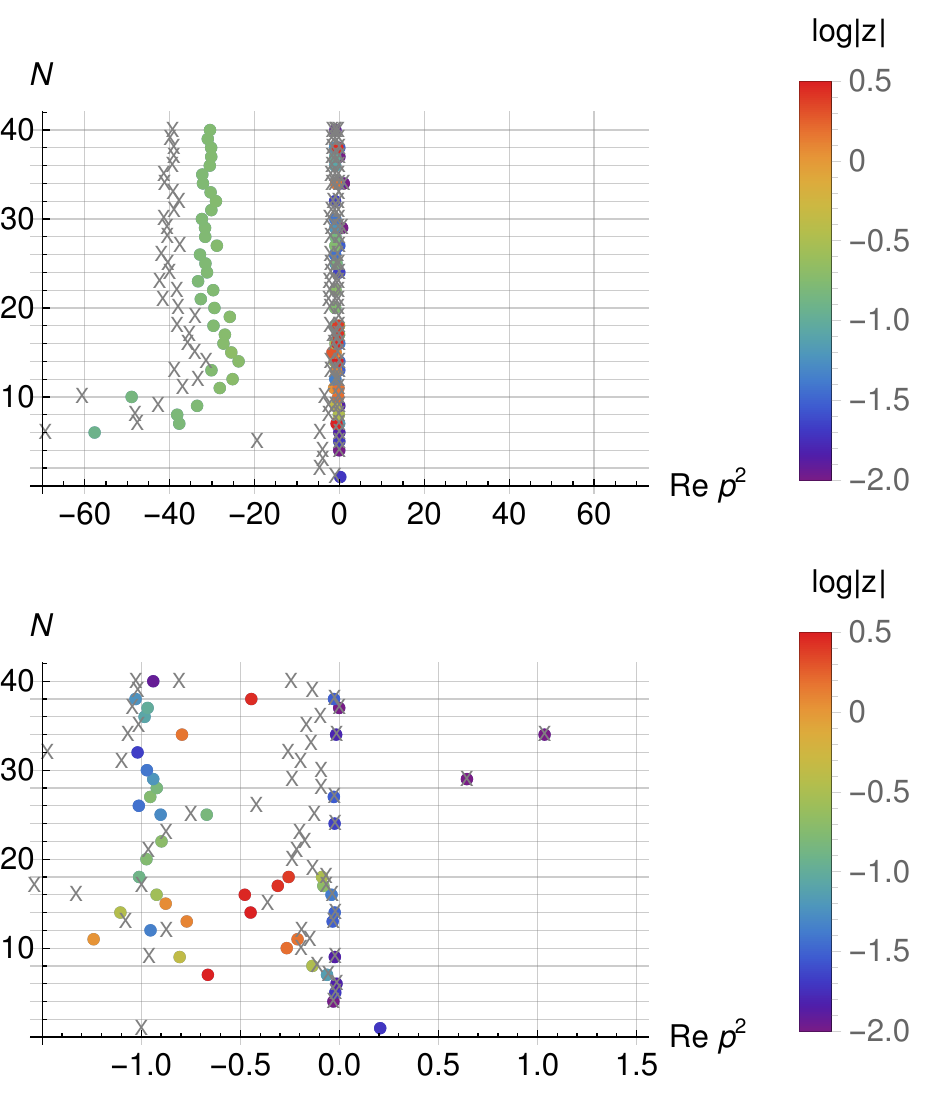}} ~
   \subfigure[$80^4$ - SA]{\includegraphics[scale=0.8]{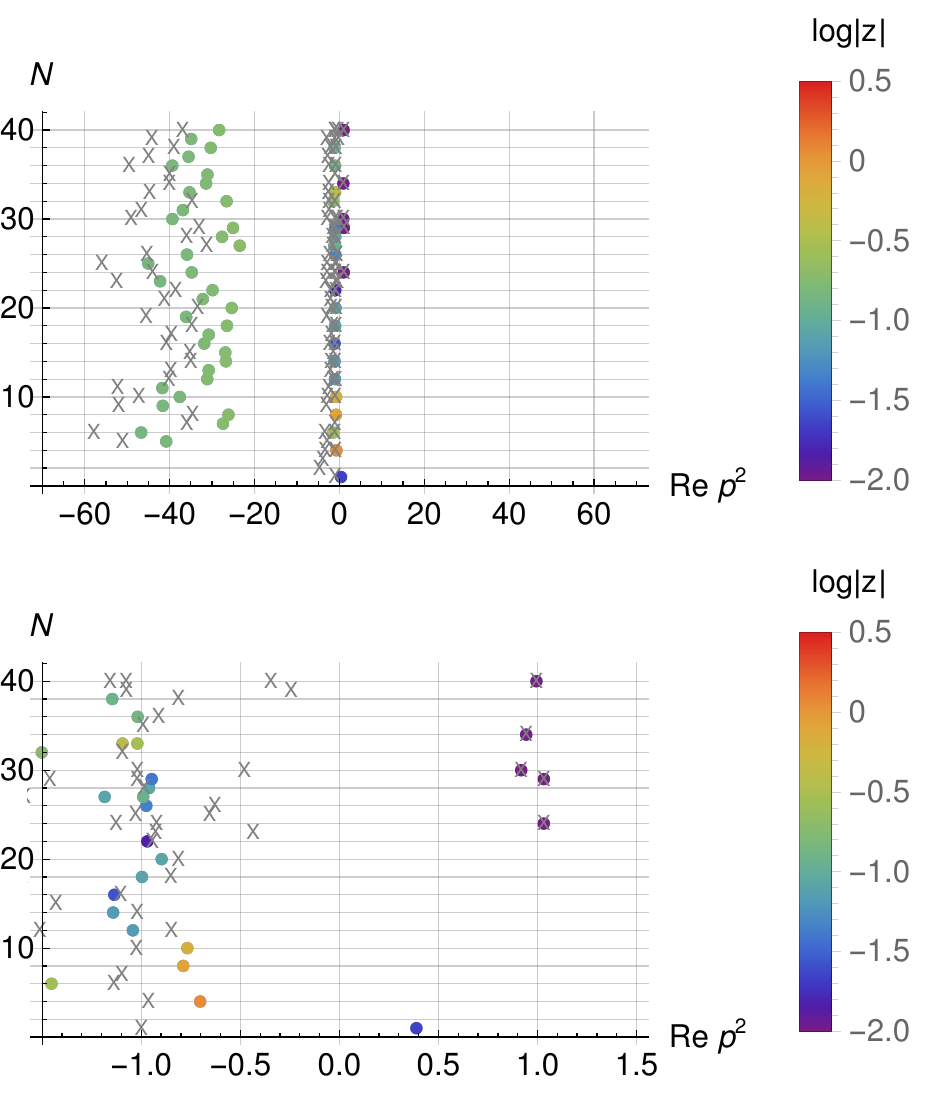}} \\
   \subfigure[$128^4$ - DE]{\includegraphics[scale=0.8]{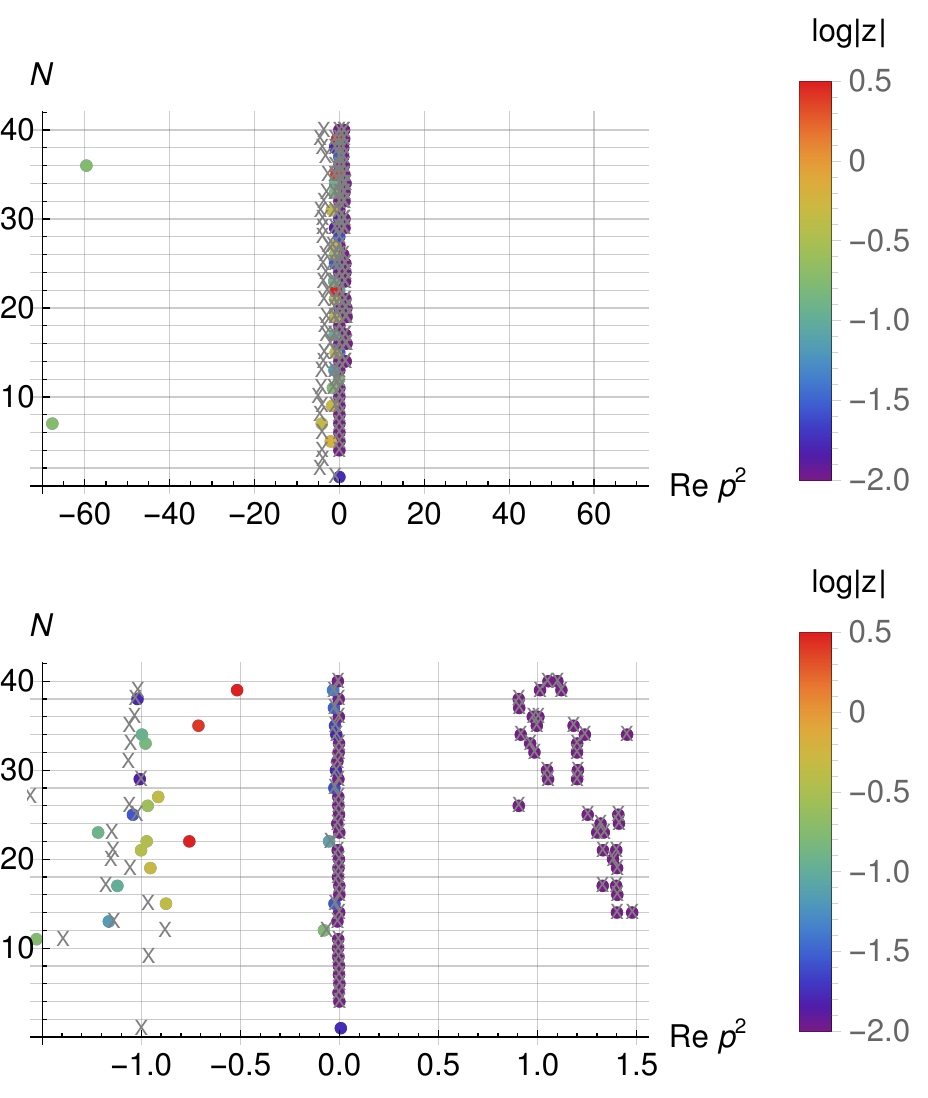}} ~
   \subfigure[$128^4$ - SA]{\includegraphics[scale=0.8]{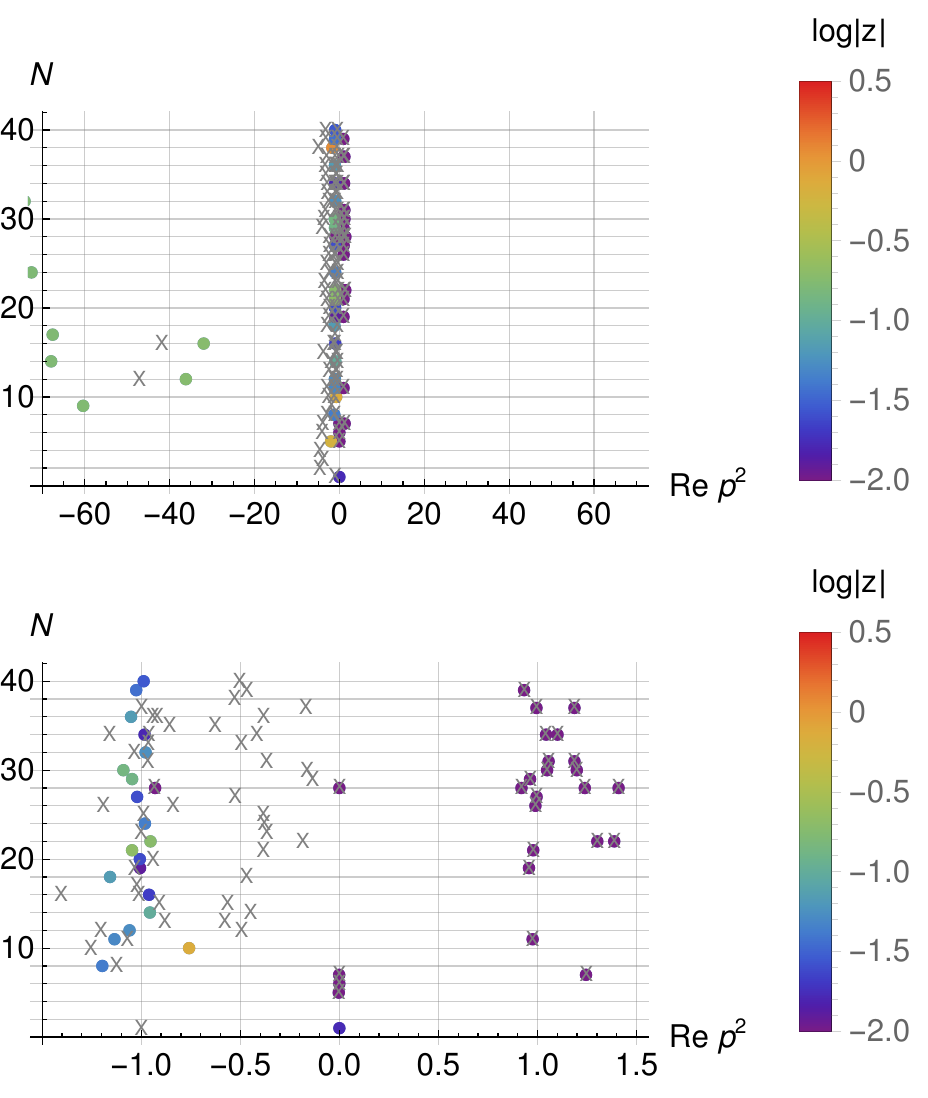}}
   \caption{The same as Fig. \ref{fig:gluon_DE_onaxis_small} for the two largest lattices.}
   \label{fig:gluon_DE_onaxis_large}
\end{figure*}

The analysis of the dominant poles of Figs. \ref{fig:gluon_DE_complex_1} to \ref{fig:gluon_SA_complex_2} suggests that the Landau gauge gluon propagator
has a pair of complex poles located around $p^2 \sim -0.3 \pm \, i \, 0.5$ GeV$^2$. Indeed, in all the Figs. there is a pole for $p^2$ around this value
with the largest absolute value for the residua. Note that, in general, the $\Im( \, p^2 )$ at the pole fluctuates significantly. 
The results using the DE method and the smallest lattice identify this pole only for the smaller $N$ and for $N \geqslant 6$ the pole is not seen anymore.
The analysis of the upper plots of Fig. \ref{fig:gluon_DE_complex_1} seems to suggest that thre is also a pole at $\Re( \, p^2 ) \geqslant 0$ that is not seen
in all the remaining simulations. The pole at $p^2 \sim -0.3 \pm \, i \, 0.5$ GeV$^2$ appears for the three largest lattices when one uses the SA method
to do the minimisation of the $\chi^2$. For the DE only for the largest lattice the pole is identified at all $N$.
This seems to suggest that the singularity associated with this momenta is connected with the infrared structure of the theory\footnote{The number of 
momenta data points considered here for the gluon propagator being 63 for the $32^4$ lattice, with 7 momenta being smaller than 1 GeV, 
126 data points for the $64^4$ lattice, with 21 momenta below 1 GeV,
168 data points for the $80^4$ lattice, with 37 momenta below 1 GeV, and 340 data points for the $128^4$ lattice, that has 131 momenta below 1 GeV.}.
One can estimate the position of the singularity looking at the dominant pole results of the largest lattice. It follows that, in all cases and for all lattices, the dominant
pole has $\Re( \, p^2 ) < 0$. If one ignores the 5 smaller and larger $N$ results,
according to the DE the singularity is at $p^2 = - (0.185 - 0.570) \pm i \, (0.301 - 0.614)$ GeV$^2$, while the SA method returns slightly smaller and looks the
singularity at $p^2 = - (0.106 - 0.308) \pm i \, (0.118 - 0.489)$ GeV$^2$. On the other hand, if takes into consideration only those $10 \leqslant N \leqslant 20$,
the DE method identify the singularity at
$p^2 = - (0.343 - 0.220) \pm i \, (0.301 - 0.546)$ GeV$^2$ and the SA returns a
$p^2 = - (0.220 - 0.150) \pm i \, (0.227 - 0.444)$ GeV$^2$.

For the gluon propagator the predictions of the Gribov-Zwanziger actions adjusted to describe the lattice data \cite{Dudal:2010tf,Cucchieri:2011ig} also suggest
the presence of complex poles that are associated with the infrared momenta. According to \cite{Dudal:2018cli} the gluon propagator has a singularity at
$p^2 = -0.268 \pm i \, 0.459$ GeV$^2$ if one uses the tree level prediction of the refined Gribov-Zwanziger action to describe the lattice data up
to $p \sim 1$ GeV. The global fits performed therein identify a pole at $p^2 = - (0.20 - 0.32)  \pm i \, (0.38 - 0.59)$ GeV$^2$.
Recall that in \cite{Dudal:2018cli} the global fits have to introduce regularisation masses and, in general, the global fits have 
$\chi^2/d.o.f.> 2$ with an exception that takes the value 1.11, whose functional form has a single pole at  $p^2 = -0.257 \pm 0.382 \, i$ GeV$^2$.
Although our current estimate points towards a pole at slightly smaller $\Re( \, p^2)$,  it is reassuring that the various estimates of the pole positions 
herein and in \cite{Dudal:2018cli} are compatible with each other.
Further, in \cite{Binosi:2019ecz} the gluon propagator was investigated with a fixed order Pad\'e approximant computed with the 
Schlessinger point method \cite{Schlessinger68}. The authors identified a pair of complex conjugate poles at $p^2 \approx -0.3 \pm i \, 0.5$ GeV$^2$ for the
same $64^4$  lattice gluon propagator data and a pole at $p^2 \approx -0.2 \pm i \, 0.35$ GeV$^2$ for the decoupling solution of the Dyson-Schwinger equations. 
Although it is difficult to make a precise comparison of the numbers, it is striking that all estimates are essentially the same and also in good agreement with the 
analysis inspired on the Gribov-Zwanziger type of actions.
A  recent analysis of the Dyson-Schwinger equations for the gluon and ghost propagators in pure Yang-Mills theory in the complex $p^2$ plan \cite{Fischer:2020xnb} 
found a singular behaviour for $p^2$ that is quite close to the complex poles given by the Pad\'e analysis.

We would like to call the readers attention that if the studies performed herein and in 
\cite{Dudal:2010tf,Oliveira:2010xc,Cucchieri:2011ig,Oliveira:2012eh,Siringo:2015wtx,Siringo:2016jrc,Dudal:2018cli,Kondo:2019rpa,Hayashi:2018giz,Binosi:2019ecz}
suggest or assume that the gluon propagator has pairs of  complex poles singularities, this is not always the case. For example,
in \cite{Strauss:2012dg} the authors solved the coupled set of Dyson-Schwinger equations for the gluon and ghost propagators, using a particular truncation,
and found no evidence of complex conjugate poles. Also the description of the massive QCD lagrangian, a particular case of the Curci-Ferrari model, investigated 
in \cite{Tissier:2010ts,Gracey:2019xom} does not point towards the presence of complex conjugate 
poles\footnote{The analysis of the pole structure of \cite{Tissier:2010ts,Gracey:2019xom} is involved and to reproduce the lattice results, the authors
perform a numerical integration using the renormalisation group improvement equations at one-loop or two-loop. 
However, taking their one-loop analytical result for the gluon propagator reproduced in their first article, 
ignoring the logarithmic corrections one obtains poles at real $p^2$.}.
Further, real valued mass gaps for the gluon and related to gluon confinement were estimated in several works
\cite{Cornwall:1981zr,Binosi:2012sj,Aguilar:2014tka,Bicudo:2015rma,Cyrol:2016tym,Huber:2020keu}.

The perturbative result for the gluon propagator has a branch cut along the real axis for negative $p^2$ and, therefore, one expects to be able to identify a branch
cut using the lattice gluon data. In Figs. \ref{fig:gluon_DE_onaxis_small} and \ref{fig:gluon_DE_onaxis_large} we show the zeros and poles for on-axis momenta
as given by the sequences of Pad\'e approximants for the different lattice data sets.
As discussed in the examples of Sec. \ref{Sec:TestFunc}, the branch cut is expected to appear as a sequence of zeros and poles. Indeed, the Figs. 
\ref{fig:gluon_DE_onaxis_small} and \ref{fig:gluon_DE_onaxis_large}  show sequences of poles and zeros along the negative real axis and close to the origin. 
However, in opposition to the results for the complex pole singularities,
the two minimisation methods do not provide consistent results when one compares the two outcomes. The DE method suggests that, if a branch cut can be
associated with the lattice data, the branch point is quite close to the origin. On the other hand, if one can read a branch cut along the negative real axis
from the analysis of the SA method, then the branch point should be at $\Re( \, p^2 ) \lesssim - 0.5$ GeV$^2$. Only the data for the largest lattice from the
SA method can suggest that the maybe-branch point can be closer to the origin. Once more, the Pad\'e analysis seems to have problems
with the exact determination of branch cuts. 
This can be either a problem of the method or that a calculation with a much larger ensemble of configurations is needed for a proper identification of
the branch cut and/or of the branch point.

One can use previous studies to estimate the window for possible values of the mass scale that regularizes the logarithm correction to the tree level perturbation
result and, in this way, estimate the branch point. Unfortunately, the reading of mass scales from other works is not straightforward and oftentimes the predictions
are for ratios of mass scales only. Despite this limitation, one can force the reading of one of the mass to be identified with the branch point. For example, relying on
the works \cite{Gracey:2019xom,Siringo:2015wtx} one can naively identify the branch point with the quoted ``gluon mass'' term that is 0.12 GeV$^2$
and 0.36 GeV$^2$, respectively. On the other hand, the  work done in  \cite{Dudal:2018cli} to fit the full set of lattice data returns a mass scale
of 0.216 GeV$^2$ \footnote{The value reported refers to the constant mass that regularizes the logarithm for infrared momenta.}. 
As stated above, one has to read these figures with great care. They all seem to be in the same ballpark and, in this sense, provide a unified picture of a set
of results obtained by rather different methods. 

In summary, our analysis suggests that the analytic structure of the gluon propagator has a pair of complex conjugate poles
together with a branch cut along the negative real axis of the Euclidean momenta. 
The corresponding branch point is located close to the origin and at the negative side of the Euclidean axis momenta.

\section{The Landau gauge ghost propagator and the Pad\'e approximants \label{Sec:ghost}}

\begin{figure}[t]
   \centering
   \includegraphics[scale=0.3]{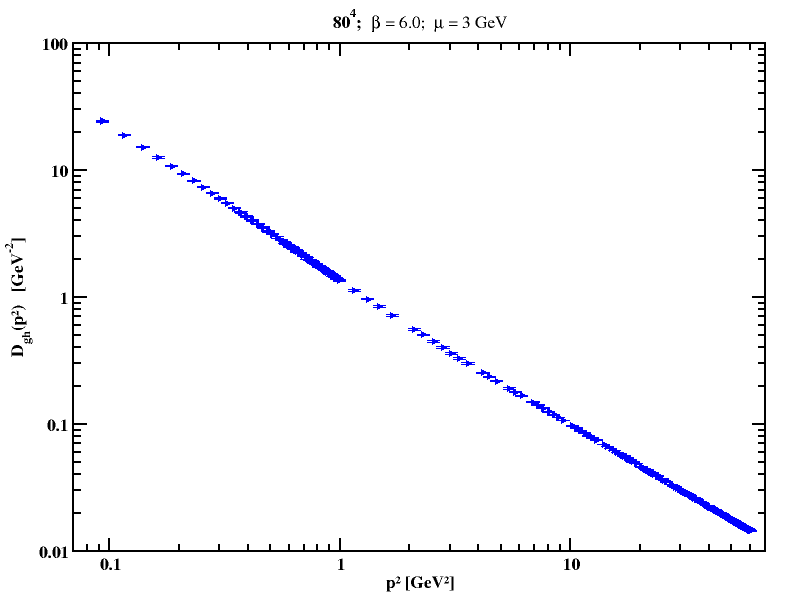} 
   \caption{The Landau gauge ghost propagator used in the Pad\'e analysis.}
   \label{fig:ghost_data}
\end{figure}

\begin{figure}[t]
   \centering
   \includegraphics[scale=0.8]{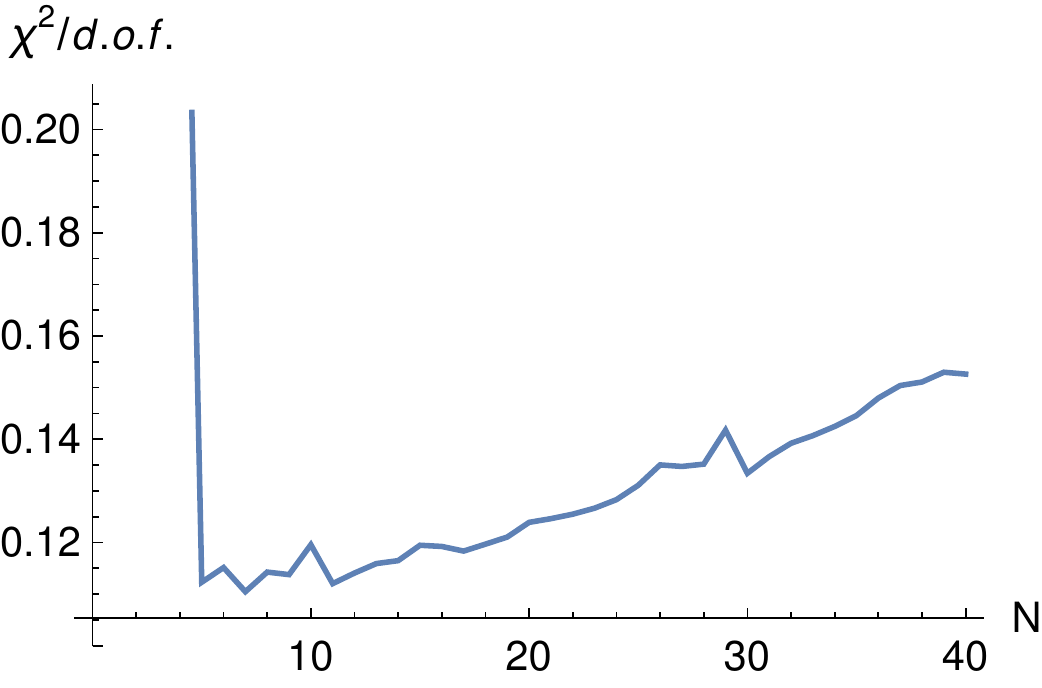} \\
   \includegraphics[scale=0.8]{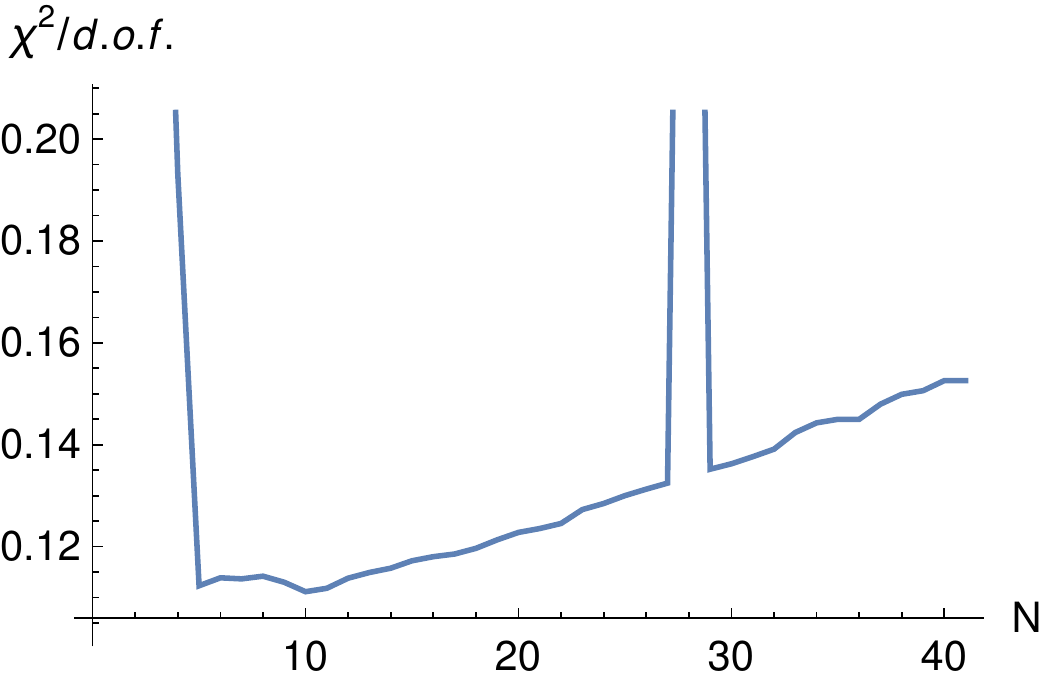}   
   \caption{Reduced $\chi^2$ at the minimum of the $\chi^2$ as obtained by the differential evolution method (top) and by the
                 simulating annealing method (bottom) as a function of the degree $N$ of the Pad\'e approximant \pade{N-1}{N}.}
   \label{fig:ghost_chi2}
\end{figure}

Let us now discuss the use of Pad\'e approximants to investigate the analytic structure of the Landau gauge ghost propagator as seen in lattice simulations.
For the lattice ghost propagator, we use the data published in \cite{Duarte:2016iko} for the simulation performed on a $80^4$ lattice with $\beta = 6.0$,
renormalised in the MOM-scheme at $\mu = 3$ GeV, as for the gluon data analysed previously. The ghost propagator lattice data can be seen in 
Fig. \ref{fig:ghost_data}. 

The reduced $\chi^2$ obtained in the minimisation of the objective function for the differential evolution and the simulated annealing
methods is reported in Fig. \ref{fig:ghost_chi2} for the \pade{N-1}{N} Pad\'e approximants. Compared to the optimal $\chi^2/d.o.f.$ obtained for the gluon
propagator that are around unit, see Fig. \ref{fig:gluon_chi2_all}, it turns out that the values of the optimal reduced $\chi^2$ for the ghost take smaller values 
and are around 0.15.

\begin{figure*}[t]
  \centering 
  \includegraphics[scale=0.8]{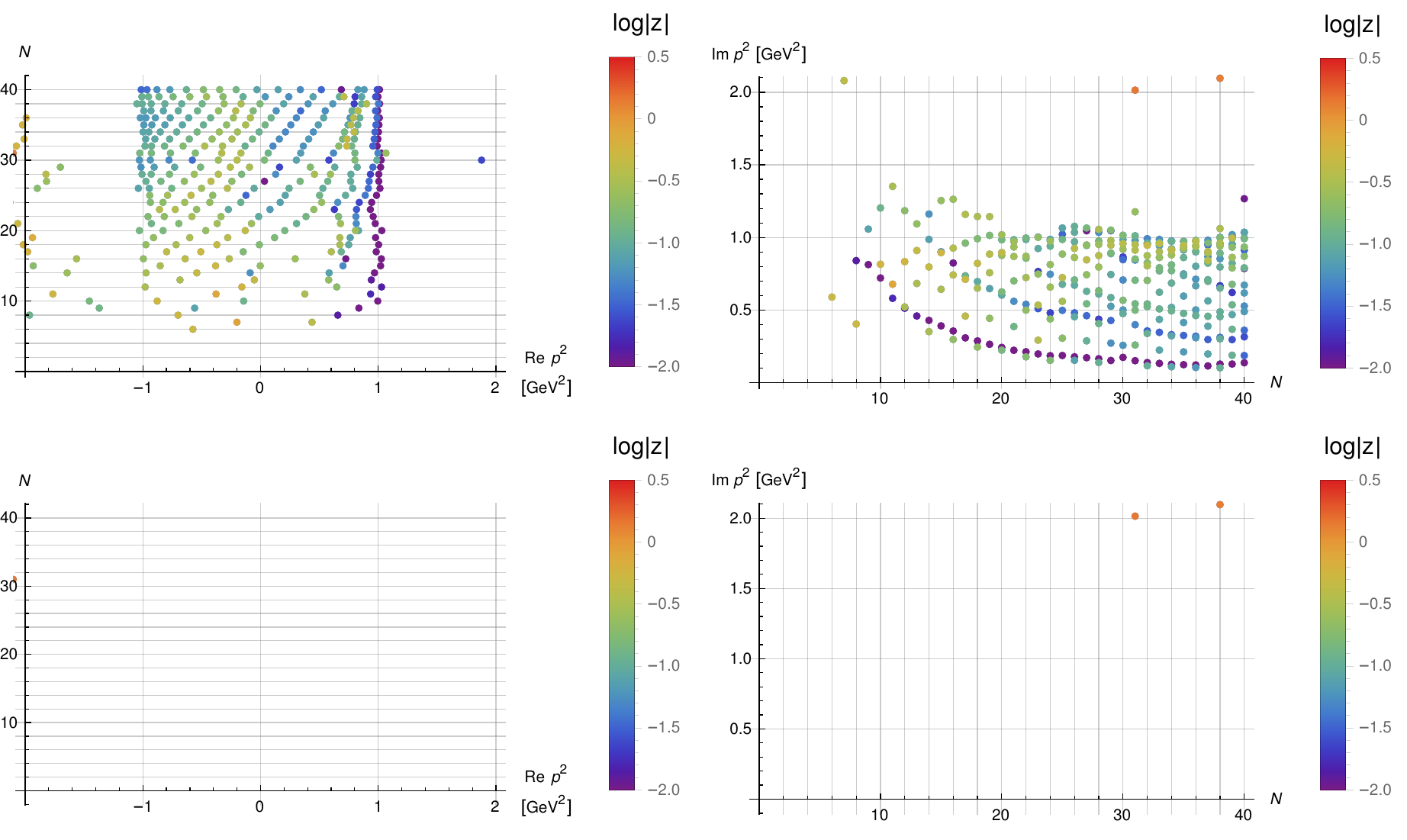} \\
   \includegraphics[scale=0.8]{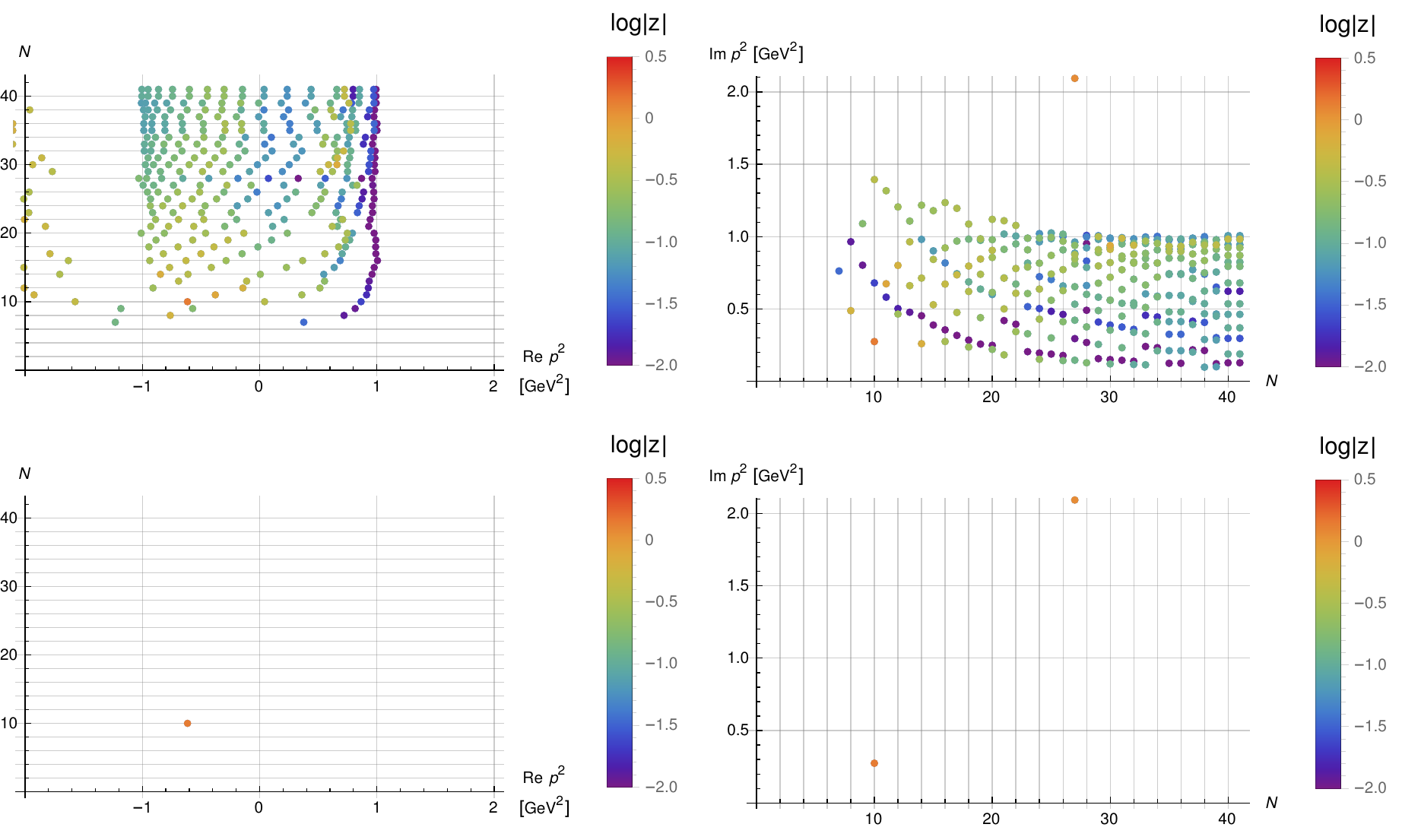}
   \caption{Evolution of the poles for complex momenta given by the Pad\'e approximants \pade{N-1}{N} and computed with the
                differential evolution method (top two plots) and the simulated annealing method (bottom two plots) for the ghost propagator. 
                For each method, the bottom plot includes only those poles that the absolute value of the residua is such that $\log |Z| > 0$.}
   \label{fig:ghost_complex_plan}
\end{figure*}

In Fig. \ref{fig:ghost_complex_plan} we show the poles for complex momenta computed from the different Pad\'e approximants  with the two minimisation
methods. As can be observed, according to the Pad\'e approximants, the ghost propagator has no complex poles.

\begin{figure}[t]
   \centering
   \includegraphics[scale=0.8]{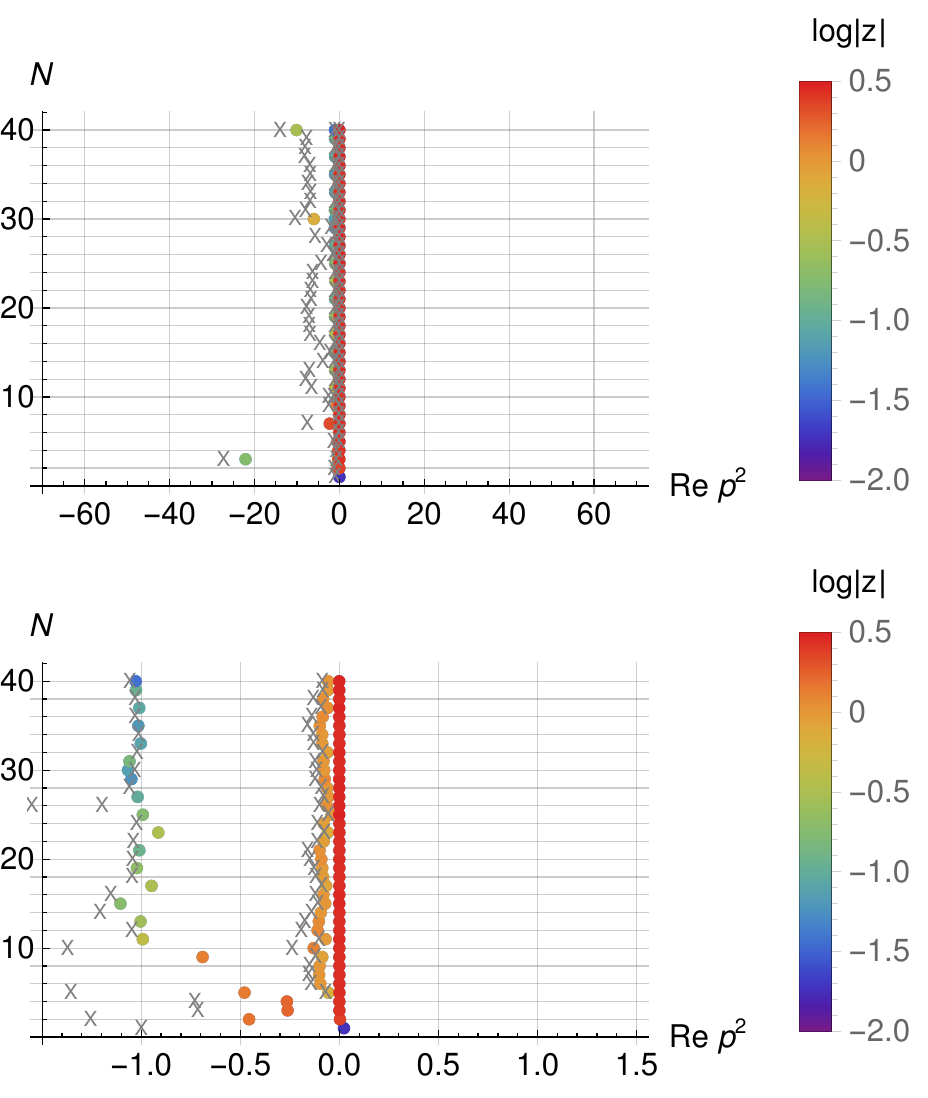} \\
   \includegraphics[scale=0.8]{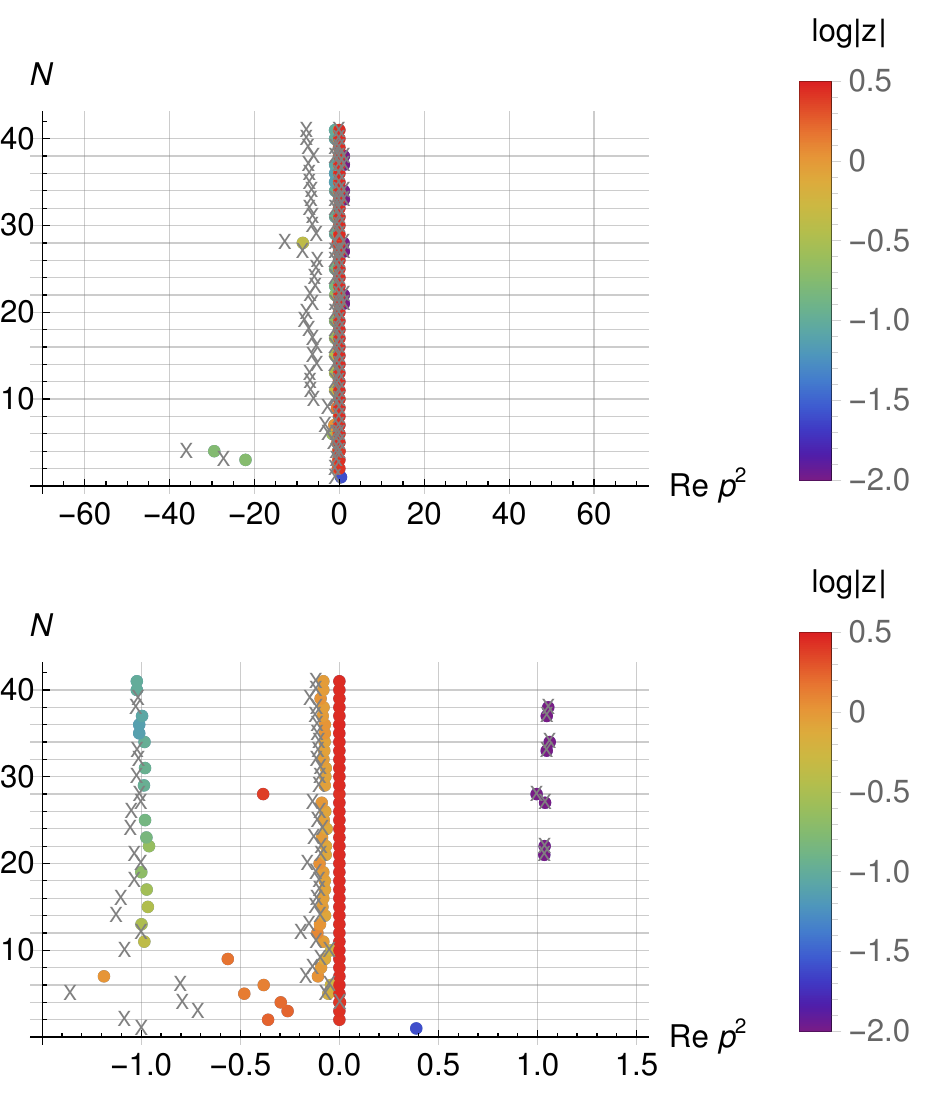}   
   \caption{Zeros (crosses) and poles (circles) computed with the Pad\'e approximants and the differential evolution method (top) and the
                 simulating annealing method (bottom) as a function of the degree $N$ for the ghost propagator. The scale on the r.h.s refers to the absolute value of the 
                 residua associated to the poles.}
   \label{fig:ghost_OnAxis}
\end{figure}

In Fig. \ref{fig:ghost_OnAxis} we resume the set of on-axis momenta poles and zeros as given by the two optimisation methods. We stress the good agreement
between the results computed with the differential evolution and the simulated annealing methods. The first remark being that, according to the Pad\'e
approximants, there is a structure of zeros and poles near the origin and towards the negative part of the Euclidean $p^2$ real axis. 
Moreover, the pole with the highest value of the absolute value of the residua is always located at $p^2 = 0$ \footnote{The exact position of the this pole
is between $p^2 = -0.001$ and $p^2 = 0.000$ for the two methods for $N > 6$.}. This is a strong indication of the presence of a pole at $p^2 = 0$, in good
agreement with the perturbative result for the ghost propagator. For negative on-axis $p^2$ and close to the origin it is observed a pole with a nearby but
not overlapping zero that, probably, is an indication of the a branch cut with a branch point located at Euclidean momenta $p^2 \sim -0.1$ GeV$^2$. 
A second sequence of poles and zeros is observed at $p^2 \sim -1$ GeV$^2$ but the residua of the poles is significantly smaller than the residua close
to the origin. The Pad\'e analysis seems to suggest that the ghost has a unique singularity located at the origin. This results is in good agreement and gives
support to the no-pole condition for the ghost propagator as proposed by Gribov \cite{Gribov:1977wm} and also supports the ghost dominance at infrared mass scales
\cite{Lerche:2002ep}.

\begin{figure}[t] 
   \centering
   \includegraphics[width=3in]{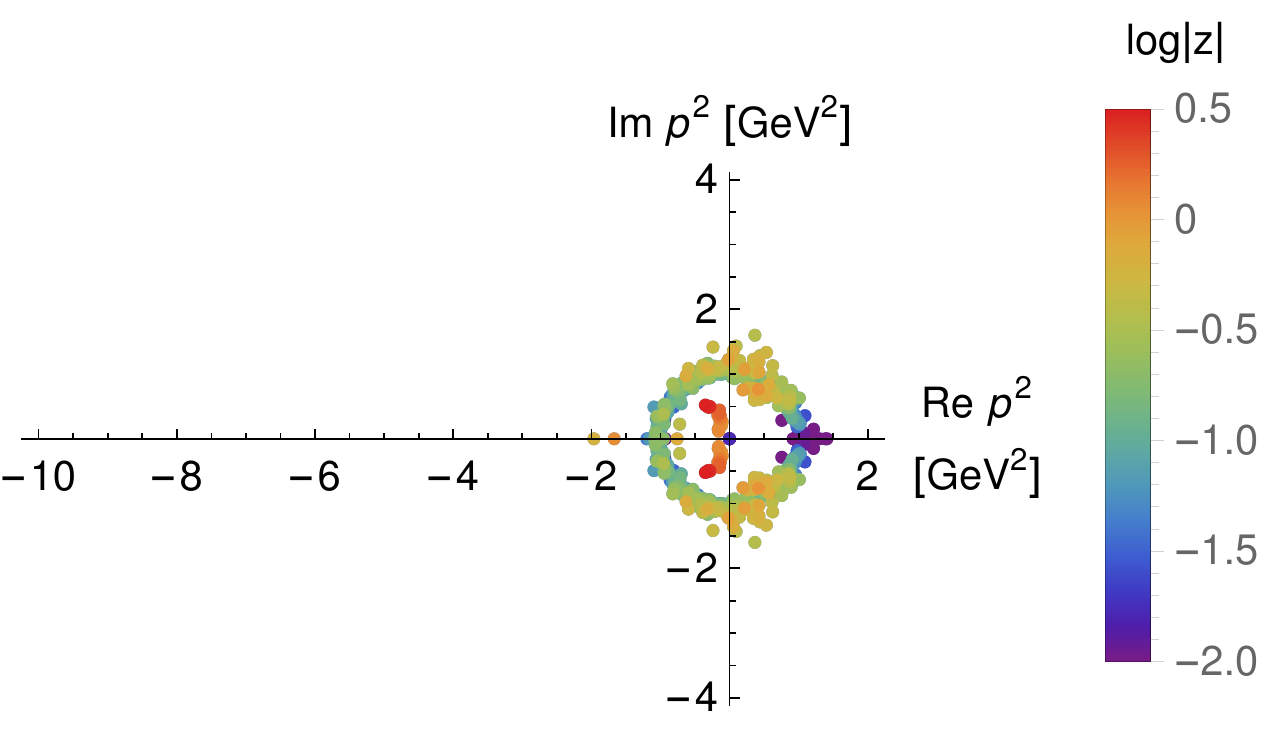} \\
   \includegraphics[width=3in]{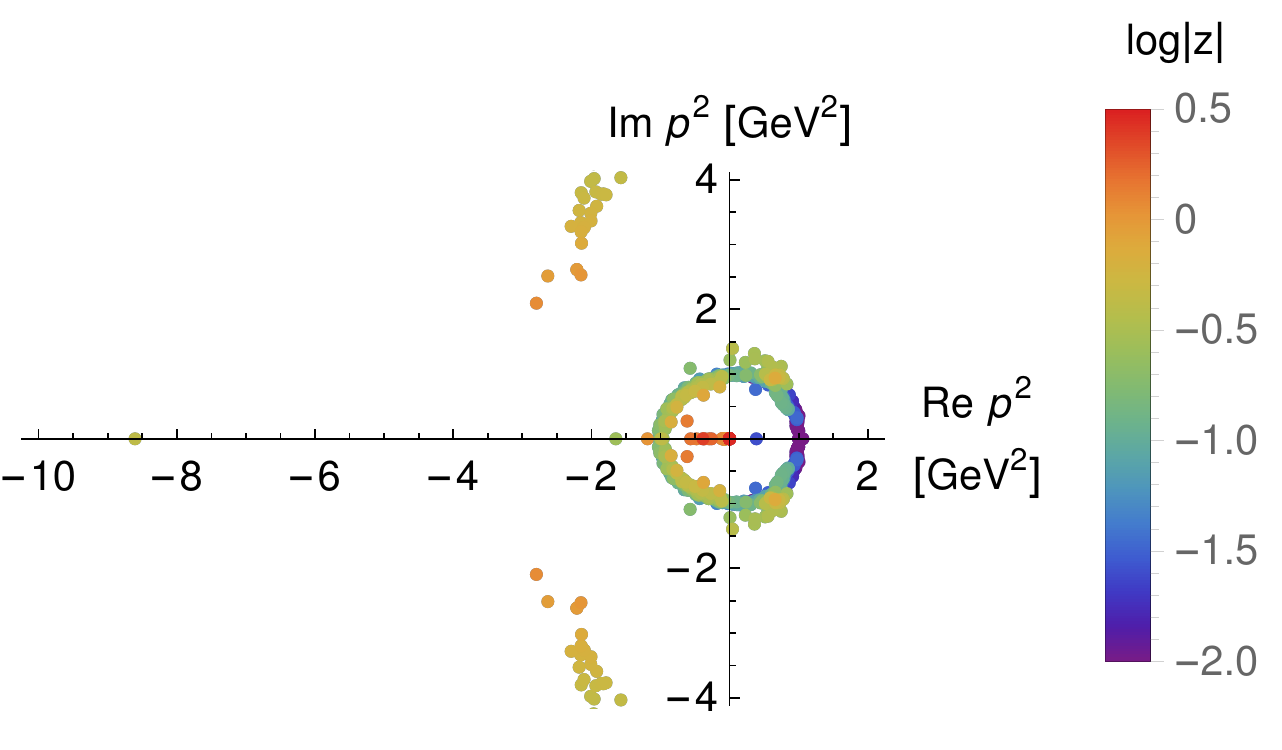}    
   \caption{Full set of poles from the Padé analysis, using the simulated annealing method,
                 for the gluon data with the $128^4$ results (top plot) and of the ghost data for the $80^4$ lattice (bottom plot) and for all $N$.}
   \label{fig:allpolesglueghost}
\end{figure}

\section{Summary and Conclusions \label{Sec:ultima}}

The access to the analytic structure of the QCD propagators is crucial if one aims to understand, for example, 
how confinement can be identified in the propagators or to compute the propagator for time-like momenta. Lattice QCD simulations
provide a first-principles calculation tool but delivers the two point correlation functions on a finite region of momenta that, typically, 
goes up to $  \lesssim 10$ GeV, on  a finite number of momenta. 
Continuum methods rely on truncations of an infinite tower of the equations and, in principle, the underlying field equations can be solved both
for real and complex momenta. Recently, modified perturbative analysis for the propagators also proved to be helpful to understand the QCD dynamics.
It is the interplay of all the methods that certainly will produce a clear picture for the propagators and sharpen our interpretation of the non-perturbative dynamics
of QCD.

Herein, we make a first try to extend the lattice data for the Landau gauge fundamental propagators of pure Yang-Mills SU(3) theory to the complex plan
and, in this way, investigate their analytic structure. We use sequences of Pad\'e approximants and look at the corresponding zeros and poles
to try to disentangle the stable poles and zeros that are translated into poles and branch cuts. From a numerical point of view, the determination of
the Pad\'e approximants is reduced to a global optimisation problem that we handle with two different methods. The patterns of the zeros and
poles given by the differential evolution and the simulated annealing methods is similar and compatible.

In the investigation of the gluon propagator a combination of several lattices is used with the aim of accessing  different ranges of momenta in order
to be able to identify  the analytic structure of this propagator.
It is for the largest physical volume that the results fluctuate less when changing the degree of the Pad\'e approximant and that the two global optimisation 
methods are closer to each other. The picture that emerges from the analysis of the different lattice data being that the gluon propagator
is described by a pair of complex conjugate poles, that are associated with the infrared momenta, together with a branch cut. 

A pair of complex conjugate poles associated with the gluon propagator is also present in other descriptions, but not all of them,
of the lattice propagator data. For example, a pair of complex conjugate poles is required by the analysis of the lattice data inspired on the 
family of Gribov-Zwanziger actions. In what concerns the location of the poles, the analysis of Sec. \ref{Sec:gluon} identifies the pole at 
$p^2 = - 0.281(62) \pm i \, 0.423 (122) $ GeV$^2$, according to the DE method, and at $p^2 = - 0.185(35) \pm i \,  0.355(108)$ GeV$^2$ for the SA method. 
The location of the complex poles predicted by the Pad\'e analysis is in good agreement with other estimates of complex poles
that can be found in the literature.

The branch cut in the gluon propagator is expected as it appears in the perturbative analysis of this two point correlation function, 
a behaviour that the lattice data should  reproduce at  higher momenta.  The Pad\'e analysis suggests a branch cut whose corresponding 
branch point is  difficult to determine, with the results of the global optimisation methods not being consistent with each other. 
The differential evolution method points towards a branch point that is close to the origin for the smallest lattices but not the largest lattice volume, 
where a structure emerges only for $\Re( \, p^2 ) \leqslant -0.5$ GeV$^2$. The simulated annealing method shows the reverse behaviour, i.e.
a structure that can be associated with a branch cut emerges at $\Re( \, p^2 ) \geqslant -0.5$ GeV$^2$ for the largest lattice.
As the Figs. \ref{fig:gluon_DE_onaxis_small} and  \ref{fig:gluon_DE_onaxis_large} show, in general, there are zeros and/or poles that can be identified with
a possible branch cut that start to appear at $\Re( \, p^2 ) \sim -0.1$ GeV$^2$ or smaller values of $\Re( \, p^2 )$. 
We take this value as an indication of a nearby branch point. 
The Pad\'e analysis is not able to provide precise information on the branch cut. This is either a limitation of the method, a limitation of a low statistical precision 
of the simulations or a combination of the two.

Our analysis of the ghost propagator is limited by the available lattice results. However, it turns out that the results associated with the ghost two point correlation
function produce a quite clear picture for the analytic structure of the propagator. 
It clearly identifies a simple pole at $p^2 = 0$, or nearby, and no further singularities are observed. 
Furthermore, the sequences of poles and zeros along the real $p^2$-axis, see Fig. \ref{fig:ghost_OnAxis}, shows a distribution that mimics what is expected 
for a branch cut. The corresponding branch point occurs at $p^2 \sim -0.1$ GeV$^2$. 
It seems that the ghost propagator is described essentially by its perturbative behaviour, i,.e.
the ghost dressing function $p^2 D_{gh}(p^2)$ has no poles but only a branch cut. 
The dressing function is finite at $p^2 = 0$ and, therefore, it seems that the non-perturbative QCD dynamics generates a mass scale that regularizes the log 
behaviour for infrared momentum. In this sense, the Pad\'e analysis for the ghost supports the no-pole condition for the ghost propagator and also the idea of
ghost dominance in the infrared region.

The analysis of the lattice data performed with the sequence of Pad\'e approximants is able to provide a picture for the analytic structure of the gluon and ghost
propagators. The problem observed with the identification of the branch cuts can, in principle, be solved by an increase on the number of gauge configurations
and on the number of momentum data points used in the calculation. 
It also would be helpful to have better control of the systematics such that the lattice simulations can provide the propagators for a larger
number of momenta. We recall that a lattice calculation of the propagators, or any Green function of the QCD fundamental fields that is not gauge invariant,
is a multiple step that starts with the sampling using a suitable gauge action and the rotation of the links towards the Landau gauge. From the computation point of view
it is the gauge fixing that is the most demanding part of the calculation. Also the computation of the ghost propagator demands solving a large set of
sparse linear systems for each gauge configuration. 
The increase of the statistical precision of the computation is feasible but certainly very time consuming. We believe that
the method explored in the current work can give us valuable information on the distribution of poles, zeros and branch cuts of the propagators
and, in principle, it can be extended for the quark propagator.

\section*{Acknowledgments}

This work was partly supported by the FCT (Portugal) Projects No. UID/FIS/04564/2019 and UID/FIS/04564/2020.
A.F.F. acknowledges financial support  from FCT (Portugal) under the project UIDB/04564/2020.
OO acknowledges Diogo Boito for calling is attention to the problem.

This work was granted access to the HPC resources of the PDC Center for High Performance Computing at the KTH Royal Institute of Technology, Sweden,
made available within the Distributed European Computing Initiative by the PRACE-2IP, receiving funding from the European Community’s Seventh Framework 
Programme (FP7/2007-2013) under grand agreement no. RI-283493. The use of Lindgren has been provided under DECI-9 project COIMBRALATT. 
We acknowledge that the results of this research have been achieved using the PRACE-3IP project (FP7 RI312763) resource Sisu based in Finland at CSC. 
The use of Sisu has been provided under DECI-12 project COIMBRALATT2. We also acknowledge the Laboratory for Advanced Computing at the
University of Coimbra (\url{http://www.uc.pt/lca}) for providing access to the HPC resource Navigator.



\end{document}